\documentclass[runningheads,orivec]{llncs}

\usepackage[T1]{fontenc}
\usepackage[utf8]{inputenc}
\usepackage{lmodern}
\usepackage{microtype}
\usepackage{etoolbox}
\usepackage{amsmath,amssymb,mathtools}
\usepackage{booktabs}
\usepackage{mathrsfs}
\usepackage{refcount}
\makeatletter
\def\caption@documentclass{standard}
\makeatother
\usepackage{subcaption}
\usepackage[ruled,vlined,linesnumbered]{algorithm2e}
\usepackage{float}
\usepackage{enumitem}
\usepackage{xcolor}
\usepackage{tikz}
\usetikzlibrary{arrows.meta,positioning,calc,fit,backgrounds,shapes.geometric}
\tikzset{
  terminal/.style={circle,fill=blue,inner sep=2pt},
  smallterminal/.style={circle,fill=blue,inner sep=1pt},
  rootvertex/.style={rectangle,fill=red,inner sep=3pt},
  steinervertex/.style={circle,fill=black,inner sep=1.4pt},
  reconnectedge/.style={red!75!black,very thick},
  pairingpath/.style={orange!90!black,ultra thick},
  markedvertex/.style={draw=orange!80!black,very thick}
}
\usepackage[colorlinks=true,linkcolor=blue!55!black,citecolor=green!45!black,urlcolor=blue!60!black,
  pdftitle={Pairwise Reconnection for Uniform Cost-Distance Steiner Trees},
  pdfsubject={Improved approximations and reconnection lower bounds}]{hyperref}
\usepackage[nameinlink,noabbrev,capitalise]{cleveref}
\crefname{equation}{equation}{equations}
\Crefname{equation}{Equation}{Equations}
\crefname{figure}{figure}{figures}
\Crefname{figure}{Figure}{Figures}
\crefname{algorithm}{algorithm}{algorithms}
\Crefname{algorithm}{Algorithm}{Algorithms}
\crefname{theorem}{theorem}{theorems}
\Crefname{theorem}{Theorem}{Theorems}
\crefname{lemma}{lemma}{lemmas}
\Crefname{lemma}{Lemma}{Lemmas}
\crefname{proposition}{proposition}{propositions}
\Crefname{proposition}{Proposition}{Propositions}
\crefname{definition}{definition}{definitions}
\Crefname{definition}{Definition}{Definitions}
\crefname{remark}{remark}{remarks}
\Crefname{remark}{Remark}{Remarks}

\spnewtheorem*{theorem*}{Theorem}{\bfseries}{\itshape}
\spnewtheorem{restatedtheorem}{Theorem}{\bfseries}{\itshape}
\newenvironment{repeatedtheorem}[1]{%
  \begingroup
  \renewcommand{\therestatedtheorem}{\getrefnumber{#1}}%
  \begin{restatedtheorem}%
}{%
  \end{restatedtheorem}%
  \endgroup
}
\DeclareMathOperator{\dist}{dist}
\DeclareMathOperator{\OPT}{OPT}
\DeclareMathOperator{\Tr}{Tr}
\newcommand{\R}{\mathbb{R}}
\newcommand{\E}{\mathbb{E}}
\newcommand{\cost}{\operatorname{cost}}
\newcommand{\Mark}{\operatorname{Mark}}
\newcommand{\Ad}{\mathcal{D}}
\newcommand{\Ap}{\mathcal{P}}
\AtEndEnvironment{proof}{\qed}
\title{Pairwise Reconnection for Uniform Cost-Distance Steiner Trees}
\subtitle{Improved Approximations and Lower Bounds}
\titlerunning{Pairwise Reconnection for Cost-Distance Steiner Trees}
\author{Edgar Perner}
\authorrunning{E. Perner}
\institute{Research Institute for Discrete Mathematics, University of Bonn\\
  \email{perner@dm.uni-bonn.de}}

\begin{document}
\maketitle

\begin{abstract}
The Uniform Cost-Distance Steiner Tree Problem asks for a rooted tree that minimizes the connection cost plus a weighted sum of root-to-terminal distances. We develop an algorithm that constructs two candidate trees: one based on direct connections to the root and another that pairs selected vertices and connects each pair to the root. The algorithm selects the tree with the smaller objective value. The resulting randomized algorithm achieves expected approximation factors of $1.698$ in the spanning case and $2.041$ in the general case, and it can be derandomized in polynomial time. The pairwise matching analysis charges part of the
augmentation cost directly to the connection and delay costs of an optimal solution, rather than only
to the separated lower bound used in previous analyses. This allows the $1.698$ factor
to fall below the maximum multiplicative gap of $1+1/\sqrt{2}$ between the optimum value and
that lower bound.

We give instance families with approximation-ratio lower bounds approaching
$1+1/\sqrt{2}$ for cut-and-reconnect algorithms that start from a minimum spanning
tree and $3/2$ for the present algorithm.
Finally, we show that the Uniform Cost-Distance Steiner Tree Problem and the Steiner-tree
version of the Cable-Trench Problem are approximation equivalent up to an arbitrarily small
additive loss.
\end{abstract}

\section{Introduction}

The \textsc{Uniform Cost-Distance Steiner Tree Problem} arises in VLSI
routing~\cite{HeldEtAl2018,DaboulEtAl2023,HeldPerner2025} and is related to network-design problems such as single-sink buy-at-bulk~\cite{AwerbuchAzar1997}. It extends the classical \textsc{Steiner Tree Problem}, where the objective is to find a minimum-length connected subgraph of an undirected graph that spans a specified subset of vertices $T$, referred to as \emph{terminals}.

In the Uniform Cost-Distance variant, the goal is to compute a rooted Steiner tree that
minimizes the sum of the tree's total edge length and a weighted sum of the distances from the
root to the terminals. The problem was introduced by Meyerson, Munagala, and
Plotkin~\cite{MeyersonMunagalaPlotkin2008}, who primarily considered the non-uniform variant, in which the total length and the distances come from separate metrics. It generalizes the \textsc{Steiner Tree Problem}
and is NP-hard even in the special case $T=V\setminus\{r\}$, where every vertex other than the root is a terminal~\cite{KhazraeiHeld2021}. In this spanning case, a minimum-length
Steiner tree is simply a minimum spanning tree and can be computed in polynomial time.
Nevertheless, the additional distance objective makes the combined problem substantially
harder: it does not admit a polynomial-time approximation scheme unless
$\mathrm{P}=\mathrm{NP}$, even when $T=V\setminus\{r\}$~\cite{BeneditoEtAl2023}.

Despite this hardness, several constant-factor approximation algorithms have been developed.
Early approaches rely on shallow-light trees, which simultaneously control the total edge
length and the root-to-terminal detours~\cite{KhullerEtAl1995}. Efficient approximation
algorithms for constructing such trees were developed in several settings, including
applications to VLSI routing~\cite{HeldRotter2013,Rotter2017,GuoZouLi2014}.

Using shallow-light spanning trees~\cite{KhullerEtAl1995}, Meyerson,
Munagala, and Plotkin~\cite{MeyersonMunagalaPlotkin2008} obtained the first constant-factor
approximation for the \textsc{Uniform Cost-Distance Steiner Tree Problem}. Later work improved the guarantee by using shallow-light Steiner
trees~\cite{HeldRotter2013,GuoZouLi2014}. By explicitly
incorporating delay weights, Khazraei and Held~\cite{KhazraeiHeld2021} and, finally, Foos, Held, and Spitzley~\cite{FoosHeldSpitzley2023Conference} reduced the approximation ratio to $1+1/\sqrt{2}<1.71$ in the
spanning case and to a ratio below $2.05$ for general terminal sets.
For general terminal sets, the best bound uses a deterministic
$(\ln(4)+\varepsilon)$-approximation for Steiner Tree as the initial tree,
for any fixed $\varepsilon>0$~\cite{ByrkaEtAl2013,TraubZenklusen2025}.

All these approaches estimate the approximation ratio by comparing the
final cost with a lower bound on the optimum that treats connection cost and delay
independently. Foos, Held, and Spitzley~\cite{FoosHeldSpitzley2023Conference} showed that the ratio between the optimal tree cost and
this lower bound approaches $1+1/\sqrt{2}$ on a family of instances. We call this the \textit{decomposition gap}.

The algorithms discussed above follow a \emph{cut-and-reconnect} structure:

\begin{definition}[Cut-and-reconnect algorithm]
A cut-and-reconnect algorithm builds an initial rooted Steiner tree, deletes an arbitrary
subset of the edges, and considers the resulting connected components. For every non-root component $K$, the algorithm chooses one
vertex $v_K\in V(K)$ and adds the edge $\{r,v_K\}$.
\end{definition}

\begin{figure}[htbp]
\centering
\begin{subfigure}[t]{0.32\textwidth}
\centering
\begin{tikzpicture}[x=0.50cm,y=0.55cm]
  \coordinate (r) at (0,0);
  \coordinate (t5) at (-2,1);
  \coordinate (t1) at (-2,3);
  \coordinate (t2) at (0,3);
  \coordinate (s) at (2,3);
  \coordinate (t3) at (4,3);
  \coordinate (t4) at (2,5);
  \draw[thick] (r)--(t5)--(t1)--(t2)--(s);
  \draw[thick] (s)--(t3) (s)--(t4);
  \node[rootvertex,label=below:$r$] at (r) {};
  \foreach \v in {t1,t2,t3,t4,t5}
    \node[terminal] at (\v) {};
  \node[steinervertex] at (s) {};
\end{tikzpicture}
\caption{Initial tree.}
\end{subfigure}\hfill
\begin{subfigure}[t]{0.32\textwidth}
\centering
\begin{tikzpicture}[x=0.50cm,y=0.55cm]
  \coordinate (r) at (0,0);
  \coordinate (t5) at (-2,1);
  \coordinate (t1) at (-2,3);
  \coordinate (t2) at (0,3);
  \coordinate (s) at (2,3);
  \coordinate (t3) at (4,3);
  \coordinate (t4) at (2,5);
  \draw[thick] (r)--(t5) (t1)--(t2) (s)--(t3) (s)--(t4);
  \node[rootvertex,label=below:$r$] at (r) {};
  \foreach \v in {t1,t2,t3,t4,t5}
    \node[terminal] at (\v) {};
  \node[steinervertex] at (s) {};
\end{tikzpicture}
\caption{Components after cutting.}
\end{subfigure}\hfill
\begin{subfigure}[t]{0.32\textwidth}
\centering
\begin{tikzpicture}[x=0.50cm,y=0.55cm]
  \coordinate (r) at (0,0);
  \coordinate (t5) at (-2,1);
  \coordinate (t1) at (-2,3);
  \coordinate (t2) at (0,3);
  \coordinate (s) at (2,3);
  \coordinate (t3) at (4,3);
  \coordinate (t4) at (2,5);
  \draw[thick] (r)--(t5) (t1)--(t2) (s)--(t3) (s)--(t4);
  \draw[reconnectedge] (r)--(t2) (r)--(s);
  \node[rootvertex,label=below:$r$] at (r) {};
  \foreach \v in {t1,t2,t3,t4,t5}
    \node[terminal] at (\v) {};
  \node[steinervertex] at (s) {};
\end{tikzpicture}
\caption{Reconnected tree.}
\end{subfigure}
\caption{Example of cut-and-reconnect. Terminals are blue, the root is red.}
\label{fig:cut-reconnect-example}
\end{figure}
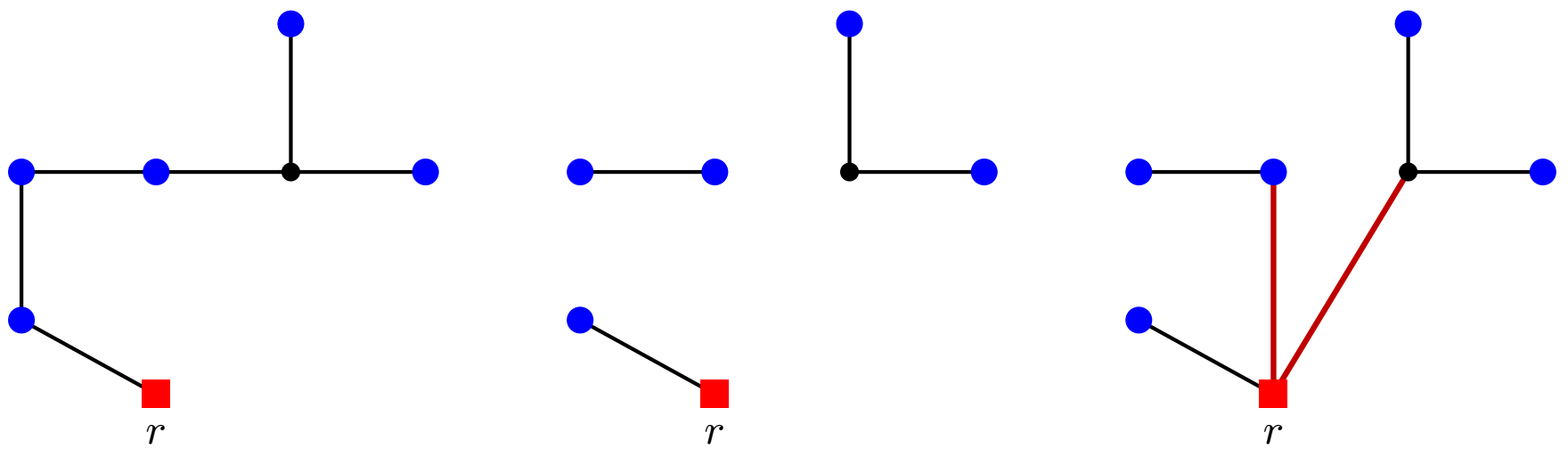

\Cref{fig:cut-reconnect-example} illustrates the procedure.

Foos, Held, and Spitzley showed in a later arXiv revision~\cite{FoosHeldSpitzley2025Arxiv} that no
cut-and-reconnect algorithm starting from a minimum-length Steiner tree can achieve an
approximation ratio better than $3/2$.

Several related special cases and generalizations of the Uniform Cost-Distance Steiner Tree
Problem have been studied. These include the
\textsc{Cable-Trench Problem}~\cite{VaskoEtAl2002}, in which all terminal weights are equal,
the two-metric Cost-Distance Problem~\cite{MeyersonMunagalaPlotkin2008}, and single-sink
Buy-at-Bulk network design~\cite{AwerbuchAzar1997}. Approximation algorithms have been
studied directly for the Steiner tree version of the Cable-Trench Problem~\cite{BeneditoEtAl2023}.

\subsection{Problem setting and notation}\label{sec:problem-setting}

\begin{definition}[Uniform Cost-Distance Steiner Tree Problem]
An instance consists of a complete undirected graph $G=(V,E)$, a pseudometric cost function
$c:E\to\R_{\ge 0}$, a root $r\in V$, a terminal set $T\subseteq V\setminus\{r\}$, and a
weight $w(t)\ge 0$ for every $t\in T$. For a connected subgraph $A$ containing
$T\cup\{r\}$, let
\[
  \cost(A)=c(A)+\sum_{t\in T}w(t)\dist_A(r,t),
  \qquad
  c(A)=\sum_{e\in E(A)}c(e),
\]
where $\dist_A(r,t)$ is the length of a shortest $r$--$t$ path in $A$ with respect to $c$.
The task is to find such a subgraph with minimum objective value $\OPT$.
\end{definition}
We call $c(A)$ the \emph{connection cost} and the weighted sum of root-to-terminal path lengths the \emph{delay cost}.
A shortest-path tree inside a solution graph $A$ has no larger connection cost than $A$ and preserves its
root distances. We may therefore assume that $A$ is a tree. We use the convention $c(v,v)=0$.

\begin{figure}[htbp]
\centering
\begin{tikzpicture}[scale=0.5]
  \coordinate (r) at (0,0);
  \coordinate (t1) at (-2,3);
  \coordinate (t2) at (0,3);
  \coordinate (s) at (2,3);
  \coordinate (t3) at (4,3);
  \coordinate (t4) at (2,5);
  \draw[thick] (r)--node[left] {$2$} (t2);
  \draw[thick] (t2)--node[below] {$1$} (t1);
  \draw[thick] (r)--node[right] {$2$} (s);
  \draw[thick] (s)--node[left] {$1$} (t4);
  \draw[thick] (s)--node[below] {$4$} (t3);
  \node[rootvertex,label=below:$r$] at (r) {};
  \node[terminal,label={[text=blue!70!black]above:$1$}] at (t1) {};
  \node[terminal,label={[text=blue!70!black]above:$1$}] at (t2) {};
  \node[terminal, inner sep = 1pt,label={[text=blue!70!black]above:$0$}] at (t3) {};
  \node[terminal, inner sep = 2.8pt,label={[text=blue!70!black]above:$2$}] at (t4) {};
  \node[steinervertex] at (s) {};
\end{tikzpicture}
\caption{A rooted tree $A$ with connection cost $c(A)=10$ and delay cost
$\textcolor{blue}{1}\cdot3+\textcolor{blue}{1}\cdot2+\textcolor{blue}{2}\cdot3+\textcolor{blue}{0}\cdot6=11$. Thus $\cost(A)=21$.}
\label{fig:cost-example}
\end{figure}
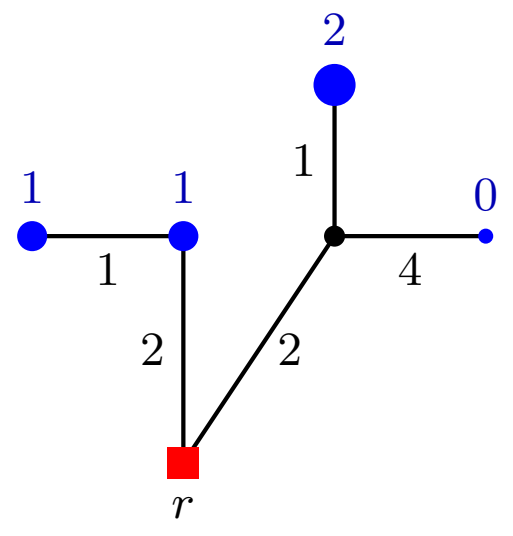

The two contributions to the objective are illustrated in
\Cref{fig:cost-example}.

Let $C_{\mathrm{ST}}$ be the cost of a minimum Steiner tree spanning $T\cup\{r\}$.
Define
\[
  D:=\sum_{t\in T}w(t)c(r,t),
  \qquad
  L:=C_{\mathrm{ST}}+D.
\]
Every feasible tree has connection cost at least $C_{\mathrm{ST}}$ and delay at least $D$.
Thus
\[
  \OPT=C^*+D^*,
  \qquad
  C^*\ge C_{\mathrm{ST}},
  \qquad
  D^*\ge D,
  \qquad
  \OPT\ge L,
\]
where $C^*$ and $D^*$ are the connection and delay costs of a fixed optimal tree $A^*$.

For $L>0$, the \textit{decomposition gap} is defined as $\OPT / L$.
Previous algorithms compare their final tree cost with $L$, so their analyses cannot beat the maximum decomposition gap.

\subsection{Our contributions}\label{sec:contributions}

\begin{enumerate}[label=\textbf{\arabic*.},leftmargin=*]

\item \textbf{Main contribution: improved approximation guarantees.}

We extend the cut-and-reconnect approach by considering two reconnection strategies in \Cref{sec:algorithm-overview}. Besides connecting selected vertices directly to the root, we introduce a second candidate that matches selected vertices and jointly connects each pair to the root through a tree. The algorithm constructs both candidates and returns the cheaper one.

We derive complementary bounds on the distance costs of the two candidates, allowing us to obtain a stronger guarantee than by considering either separately. Furthermore, we bound the matching cost of the pair candidate using edge-disjoint paths in an optimal Cost-Distance Steiner tree, charging part of the reconnection cost directly to the optimum rather than only to the lower bound $L$. Together, these ingredients yield the first approximation guarantee below the maximum decomposition gap.

We establish a parameterized approximation guarantee in \Cref{thm:general-factor}, depending on the Steiner tree approximation factor $\omega$ and a parameter $\lambda$. Explicit evaluation in \Cref{sec:one-dimensional-certification} yields approximation factors of $1.698$ in the spanning case and $2.041$ in the general case, improving the previous guarantees of $1.7071$ and $2.0478$ by Foos, Held, and Spitzley~\cite{FoosHeldSpitzley2023Conference}.

\item \textbf{Lower bounds for reconnection algorithms.}
\begin{repeatedtheorem}{thm:reconnection-lb}[Reconnection lower bounds]
For every $\varepsilon>0$, there are spanning instances in the $\ell_1$ metric such that
\begin{enumerate}[label=(\roman*),leftmargin=*]
  \item cut-and-reconnect algorithms that start with a minimum spanning tree have an approximation ratio of at least
        $1+1/\sqrt{2}-\varepsilon$;
  \item the algorithm presented in this paper has a worst-case approximation ratio of at least
        $3/2-\varepsilon$.
\end{enumerate}
\end{repeatedtheorem}
This class of cut-and-reconnect algorithms includes the algorithm of Foos, Held, and
Spitzley. Part~(i) closes the gap to their approximation guarantee
for the spanning case and shows that direct reconnection alone is insufficient to
achieve a ratio below the decomposition gap.

\item \textbf{Relation to the Cable-Trench Problem.}

We clarify the relation between the Uniform Cost-Distance Steiner Tree Problem and the Cable-Trench Problem. The Steiner tree version of the Cable-Trench Problem is the special case of the Uniform Cost-Distance Steiner Tree Problem where all terminals have equal weight~\cite{BeneditoEtAl2023}. To our knowledge, no better guarantees are known for that special case.

We show that the two problems are in fact approximation equivalent up to an
arbitrarily small additive loss.
\begin{repeatedtheorem}{thm:cable-reduction}[Reduction to Cable-Trench with Steiner points]
Suppose that the Steiner tree version of the Cable-Trench Problem has an $\alpha$-approximation. Then, for
every fixed $\varepsilon>0$, the Uniform Cost-Distance Steiner Tree Problem has an
$(\alpha+\varepsilon)$-approximation. For fixed $\alpha$, the running time is polynomial in the input size and in
$1/\varepsilon$.
\end{repeatedtheorem}

\end{enumerate}

We first explain the new algorithm and prove the parameterized approximation guarantee,
then establish the lower bounds and the reduction to the Cable-Trench Problem.

\section{Algorithmic overview}\label{sec:algorithm-overview}

In \Cref{alg:shifted-gates} we construct two trees $\Ad$ and $\Ap$ based on an initial Steiner tree $A_0$ and return the cheaper of the two. Similarly to the cut-and-reconnect framework, $A_0$ is an approximation of a minimum-length Steiner tree, and the direct candidate $\Ad$ is obtained by adding root-to-terminal connections to $A_0$.

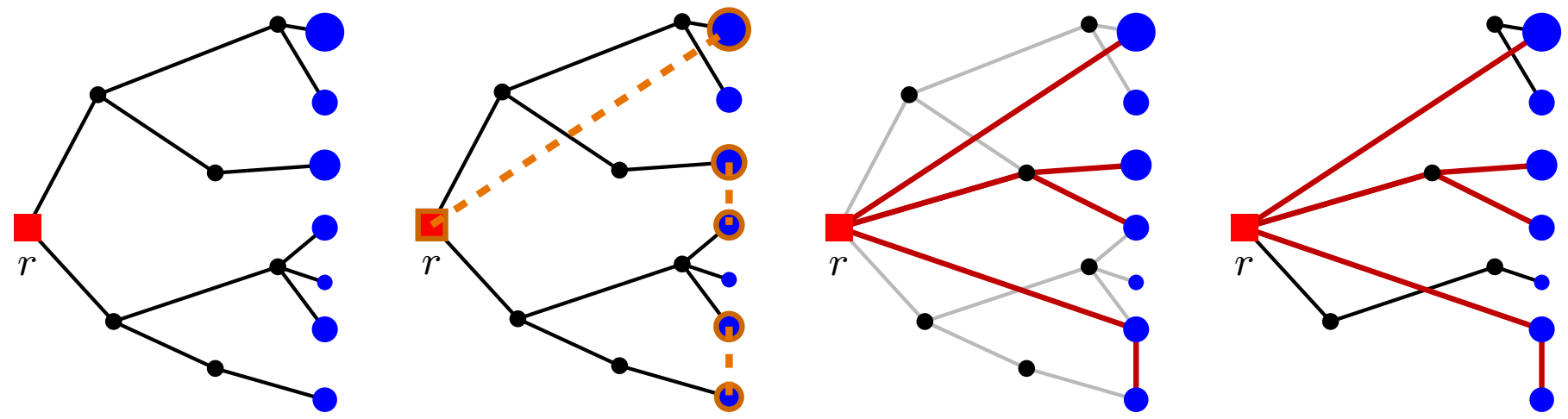
\begin{figure}[H]
\centering
\begin{subfigure}[t]{0.235\textwidth}
\centering
\begin{tikzpicture}[
    x=0.6cm,y=0.6cm,
    gate/.style={circle,draw=orange!80!black,fill=orange!30,inner sep=1.7pt},
    steiner/.style={circle,fill=black,inner sep=1.3pt}
  ]
  \coordinate (r)  at (0,0);
  \coordinate (a)  at (0.9,1.7);
  \coordinate (b)  at (1.1,-1.2);
  \coordinate (c)  at (3.2,2.6);
  \coordinate (d)  at (2.4,0.7);
  \coordinate (e)  at (3.2,-0.5);
  \coordinate (f)  at (2.4,-1.8);
  \coordinate (t1) at (3.8,2.5);
  \coordinate (t2) at (3.8,1.6);
  \coordinate (t3) at (3.8,0.8);
  \coordinate (t4) at (3.8,0.0);
  \coordinate (t5) at (3.8,-0.7);
  \coordinate (t6) at (3.8,-1.3);
  \coordinate (t7) at (3.8,-2.2);

  \draw[thick] (r)--(a)--(c);
  \draw[thick] (a)--(d);
  \draw[thick] (r)--(b)--(f);
  \draw[thick] (b)--(e);
  \draw[thick] (c)--(t1);
  \draw[thick] (c)--(t2);
  \draw[thick] (d)--(t3);
  \draw[thick] (e)--(t4);
  \draw[thick] (e)--(t5);
  \draw[thick] (e)--(t6);
  \draw[thick] (f)--(t7);

  \node[rootvertex,label=below:$r$] at (r) {};
  \node[steiner] at (a) {};
  \node[steiner] at (b) {};
  \node[steiner] at (c) {};
  \node[steiner] at (d) {};
  \node[steiner] at (e) {};
  \node[steiner] at (f) {};

  \node[terminal, inner sep = 3pt] at (t1) {};
  \node[terminal] at (t2) {};
  \node[terminal, inner sep = 2.4pt] at (t3) {};
  \node[terminal] at (t4) {};
  \node[terminal, inner sep = 1.2pt] at (t5) {};
  \node[terminal] at (t6) {};
  \node[terminal, inner sep = 1.9pt] at (t7) {};
\end{tikzpicture}
\caption{Initial tree $A_0$.}
\end{subfigure}\hfill
\begin{subfigure}[t]{0.235\textwidth}
\centering
\begin{tikzpicture}[
    x=0.6cm,y=0.6cm,
    gate/.style={circle,draw=orange!80!black,fill=orange!30,inner sep=1.7pt},
    steiner/.style={circle,fill=black,inner sep=1.3pt}
  ]
  \coordinate (r)  at (0,0);
  \coordinate (a)  at (0.9,1.7);
  \coordinate (b)  at (1.1,-1.2);
  \coordinate (c)  at (3.2,2.6);
  \coordinate (d)  at (2.4,0.7);
  \coordinate (e)  at (3.2,-0.5);
  \coordinate (f)  at (2.4,-1.8);
  \coordinate (t1) at (3.8,2.5);
  \coordinate (t2) at (3.8,1.6);
  \coordinate (t3) at (3.8,0.8);
  \coordinate (t4) at (3.8,0.0);
  \coordinate (t5) at (3.8,-0.7);
  \coordinate (t6) at (3.8,-1.3);
  \coordinate (t7) at (3.8,-2.2);

  \draw[thick] (r)--(a)--(c);
  \draw[thick] (a)--(d);
  \draw[thick] (r)--(b)--(f);
  \draw[thick] (b)--(e);
  \draw[thick] (c)--(t1);
  \draw[thick] (c)--(t2);
  \draw[thick] (d)--(t3);
  \draw[thick] (e)--(t4);
  \draw[thick] (e)--(t5);
  \draw[thick] (e)--(t6);
  \draw[thick] (f)--(t7);

  \node[draw=orange!80!black,very thick,rootvertex,label=below:$r$] at (r) {};
  \node[steiner] at (a) {};
  \node[steiner] at (b) {};
  \node[steiner] at (c) {};
  \node[steiner] at (d) {};
  \node[steiner] at (e) {};
  \node[steiner] at (f) {};

  \node[draw=orange!80!black,very thick, terminal, inner sep = 3pt] at (t1) {};
  \node[terminal] at (t2) {};
  \node[draw=orange!80!black,very thick,terminal, inner sep = 2.4pt] at (t3) {};
  \node[draw=orange!80!black,very thick,terminal] at (t4) {};
  \node[terminal, inner sep = 1.2pt] at (t5) {};
  \node[draw=orange!80!black,very thick,terminal] at (t6) {};
  \node[draw=orange!80!black,very thick,terminal, inner sep = 1.9pt] at (t7) {};

  \draw[dashed,orange!90!black,ultra thick] (r) to (t1);
  \draw[dashed,orange!90!black,ultra thick] (t3) to (t4);
  \draw[dashed,orange!90!black,ultra thick] (t6) to (t7);
\end{tikzpicture}
\caption{Minimum-cost matching.}
\end{subfigure}\hfill
\begin{subfigure}[t]{0.235\textwidth}
\centering
\begin{tikzpicture}[
    x=0.6cm,y=0.6cm,
    gate/.style={circle,draw=orange!80!black,fill=orange!30,inner sep=1.7pt},
    steiner/.style={circle,fill=black,inner sep=1.3pt}
  ]
  \coordinate (r)  at (0,0);
  \coordinate (a)  at (0.9,1.7);
  \coordinate (b)  at (1.1,-1.2);
  \coordinate (c)  at (3.2,2.6);
  \coordinate (d)  at (2.4,0.7);
  \coordinate (e)  at (3.2,-0.5);
  \coordinate (f)  at (2.4,-1.8);
  \coordinate (t1) at (3.8,2.5);
  \coordinate (t2) at (3.8,1.6);
  \coordinate (t3) at (3.8,0.8);
  \coordinate (t4) at (3.8,0.0);
  \coordinate (t5) at (3.8,-0.7);
  \coordinate (t6) at (3.8,-1.3);
  \coordinate (t7) at (3.8,-2.2);

  \draw[thick,gray!55] (r)--(a)--(c);
  \draw[thick,gray!55] (a)--(d);
  \draw[thick,gray!55] (r)--(b)--(f);
  \draw[thick,gray!55] (b)--(e);
  \draw[thick,gray!55] (c)--(t1);
  \draw[thick,gray!55] (c)--(t2);
  \draw[thick,gray!55] (d)--(t3);
  \draw[thick,gray!55] (e)--(t4);
  \draw[thick,gray!55] (e)--(t5);
  \draw[thick,gray!55] (e)--(t6);
  \draw[thick,gray!55] (f)--(t7);

  \draw[red!75!black,very thick] (r)--(t1);
  \draw[red!75!black,very thick] (r)--(d)--(t3);
  \draw[red!75!black,very thick] (r)--(d)--(t4);
  \draw[red!75!black,very thick] (r)--(t6)--(t7);

  \node[rootvertex,label=below:$r$] at (r) {};
  \node[steiner] at (a) {};
  \node[steiner] at (b) {};
  \node[steiner] at (c) {};
  \node[steiner] at (d) {};
  \node[steiner] at (e) {};
  \node[steiner] at (f) {};

  \node[terminal, inner sep = 3pt] at (t1) {};
  \node[terminal] at (t2) {};
  \node[terminal, inner sep = 2.4pt] at (t3) {};
  \node[terminal] at (t4) {};
  \node[terminal, inner sep = 1.2pt] at (t5) {};
  \node[terminal] at (t6) {};
  \node[terminal, inner sep = 1.9pt] at (t7) {};
\end{tikzpicture}
\caption{Added pair trees $F_{u,v}$.}
\end{subfigure}\hfill
\begin{subfigure}[t]{0.235\textwidth}
\centering
\begin{tikzpicture}[
    x=0.6cm,y=0.6cm,
    steiner/.style={circle,fill=black,inner sep=1.3pt}
  ]
  \coordinate (r)  at (0,0);
  \coordinate (a)  at (0.9,1.7);
  \coordinate (b)  at (1.1,-1.2);
  \coordinate (c)  at (3.2,2.6);
  \coordinate (d)  at (2.4,0.7);
  \coordinate (e)  at (3.2,-0.5);
  \coordinate (f)  at (2.4,-1.8);
  \coordinate (t1) at (3.8,2.5);
  \coordinate (t2) at (3.8,1.6);
  \coordinate (t3) at (3.8,0.8);
  \coordinate (t4) at (3.8,0.0);
  \coordinate (t5) at (3.8,-0.7);
  \coordinate (t6) at (3.8,-1.3);
  \coordinate (t7) at (3.8,-2.2);

  \draw[thick] (t1)--(c)--(t2);
  \draw[thick] (r)--(b)--(e)--(t5);
  \draw[red!75!black,very thick] (r)--(t1);
  \draw[red!75!black,very thick] (r)--(d)--(t3);
  \draw[red!75!black,very thick] (r)--(d)--(t4);
  \draw[red!75!black,very thick] (r)--(t6)--(t7);

  \node[rootvertex,label=below:$r$] at (r) {};
  \node[steiner] at (b) {};
  \node[steiner] at (c) {};
  \node[steiner] at (e) {};
  \node[steiner] at (d) {};
  \node[terminal, inner sep = 3pt] at (t1) {};
  \node[terminal] at (t2) {};
  \node[terminal, inner sep = 2.4pt] at (t3) {};
  \node[terminal] at (t4) {};
  \node[terminal, inner sep = 1.2pt] at (t5) {};
  \node[terminal] at (t6) {};
  \node[terminal, inner sep = 1.9pt] at (t7) {};
\end{tikzpicture}
\caption{Shortest-path tree of the union.}
\end{subfigure}
\caption{Example of the construction of $\Ap$. Orange outlines denote marked vertices.}
\label{fig:pair-reconnection-example}
\end{figure}

The pair candidate $\Ap$ is obtained from $A_0$ by finding a minimum-cost matching of marked vertices and connecting each matched pair $u,v$ with an $\{r,u,v\}$-Cost-Distance Steiner tree that is optimal with respect to modified weights.
The construction is displayed in \Cref{fig:pair-reconnection-example}.

We explain how vertices are marked for direct or pairwise reconnection in the next subsection. Given $q>0$, we call this set $\Mark_q(\theta)$. The input parameter $\lambda$ in \Cref{alg:shifted-gates} indirectly controls the number of marked vertices.

\begin{algorithm}[H]
\small
\caption{Direct and pairwise reconnection}
\label{alg:shifted-gates}
\DontPrintSemicolon
\SetKwInput{KwIn}{Input}
\SetKwInput{KwOut}{Output}

\KwIn{Metric Uniform Cost-Distance instance, Steiner approximation factor $\omega$, a parameter $\lambda>0$}
\KwOut{Feasible rooted tree}

Compute an $\omega$-approximation $A_0$ to the shortest Steiner tree on $T \cup \{r\}$ with respect to $c$\;
\If{$c(A_0)=0$}{
    \Return{$A_0$}\;
}
Set the delay bound $D \gets \sum_{t\in T}w(t)c(r,t)$\;
\If{$D=0$}{
    \Return{\textup{a shortest-path tree rooted at }$r$ \textup{in} $A_0\cup\{\{r,t\}:w(t)>0\}$}\;
}
Set $\mu\gets\lambda\sqrt{2D/c(A_0)}$\;
\medskip
\textbf{Direct reconnection:}\;
Sample $\theta \in [0,\mu)$ uniformly and select vertices $\Mark_{\mu}(\theta) \subseteq T \cup \{r\}$ for direct reconnection as in \Cref{sec:gate_pos}\;
$\Ad_{\mu}(\theta) \gets \text{shortest-path tree rooted at } r
    \text{ in } A_0 \cup \bigl\{\{r,t\} : t \in \Mark_{\mu}(\theta)\setminus\{r\}\bigr\}$\;

\medskip
\textbf{Pairwise reconnection:}\;
Sample $\theta' \in [0,\mu/2)$ uniformly, select vertices $\Mark_{\mu/2}(\theta') \subseteq T \cup \{r\}$ for pairwise reconnection and compute modified weights $w'$ as in \Cref{sec:gate_pos}\;
\ForEach{$\{u,v\} \subseteq \Mark_{\mu/2}(\theta'),\ u \neq v$}{
    Compute an optimal Cost-Distance Steiner tree $F_{u,v}$ for root $r$, terminals $\{u,v\}\setminus \{r\}$, and weights $w'$ according to \Cref{lem:two-terminal}. Let
    $K_{u,v}$ be the corresponding cost\;
}
$M \gets \text{minimum-cost perfect matching on } \Mark_{\mu/2}(\theta')
    \text{ with respect to } K$\;
$\Ap_{\mu/2}(\theta') \gets \text{shortest-path tree rooted at } r
    \text{ in } A_0 \cup \displaystyle\bigcup_{\{u,v\} \in M} F_{u,v}$\;
\medskip
\Return{$\displaystyle\arg\min_{A \in \{\Ad_{\mu}(\theta),\Ap_{\mu/2}(\theta')\}}
    \operatorname{cost}(A)$}\;
\end{algorithm}

\subsection{Determining the marked vertices via gates}

We now explain how the set $\Mark_q(\theta)$ of marked
vertices is constructed. The idea is to
randomly discretize the terminal weights into multiples
of a common value $q>0$, which we call the \emph{capacity}, while preserving each
weight in expectation. A terminal is added to $\Mark_q(\theta)$ if its
discretized weight is positive. If the number of such
terminals is odd, we additionally mark the root, ensuring
that $\Mark_q(\theta)$ has even cardinality as required
for pairwise reconnection.

We represent each unit of discretized weight by a
\emph{gate} of capacity $q$. Each gate is assigned
to a vertex in $T \cup \{r\}$, called its \emph{anchor}.
Several gates may share the same anchor, but each
terminal is marked only once.

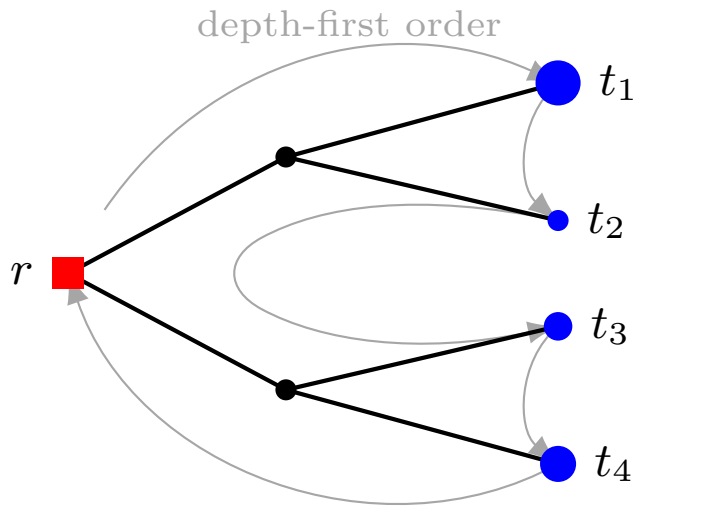
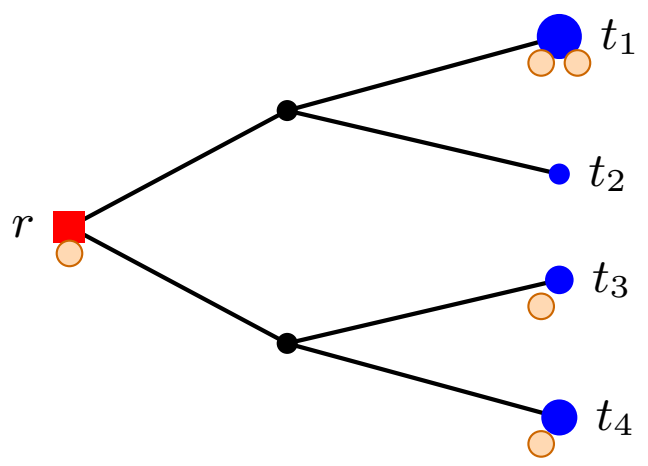
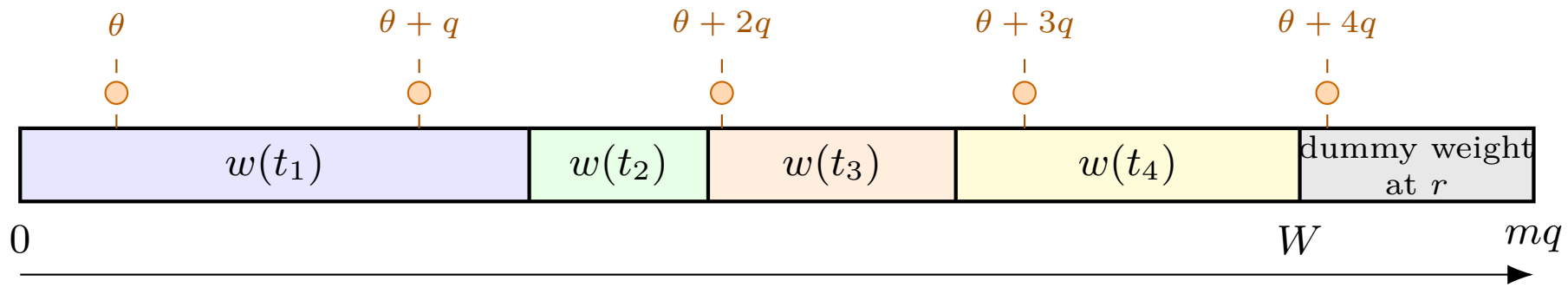
\begin{figure}[H]
\centering
\begin{subfigure}[t]{0.49\textwidth}
\centering
\begin{tikzpicture}[x=1.2cm,y=0.7cm, gate/.style={circle,draw=orange!80!black,fill=orange!30,inner sep=1.7pt}]
  \coordinate (r) at (0,0);
  \coordinate (a) at (1.2,1.1);
  \coordinate (b) at (1.2,-1.1);
  \coordinate (t1) at (2.7,1.8);
  \coordinate (t2) at (2.7,0.5);
  \coordinate (t3) at (2.7,-0.5);
  \coordinate (t4) at (2.7,-1.8);
  \draw[-{Latex[length=2mm]},gray!70] (0.2, 0.6) to[bend left=40] (t1);
  \draw[-{Latex[length=2mm]},gray!70] (t1) to[bend right=50] (t2);
  \draw[-{Latex[length=2mm]},gray!70, rounded corners=5mm] (t2) to[bend right=20] (0.7,0) to[bend right=20] (t3);
  \draw[-{Latex[length=2mm]},gray!70] (t3) to[bend right=50] (t4);
  \draw[-{Latex[length=2mm]},gray!70] (t4) to[bend left=50] (r);
  \draw[thick] (r)--(a)--(t1);
  \draw[thick] (a)--(t2);
  \draw[thick] (r)--(b)--(t3);
  \draw[thick] (b)--(t4);

  \node[rootvertex,label=left:$r$] at (r) {};
  \node[steinervertex] at (a) {};
  \node[steinervertex] at (b) {};
  \node[terminal,label=right:$t_1$,inner sep=3pt] at (t1) {};
  \node[terminal,label=right:$t_2$,inner sep=1.4pt] at (t2) {};
  \node[terminal,label=right:$t_3$,inner sep=1.9pt] at (t3) {};
  \node[terminal,label=right:$t_4$,inner sep=2.4pt] at (t4) {};
  \node[font=\scriptsize,gray!70] at (1.55,2.33) {depth-first order};
\end{tikzpicture}
\caption{Initial tree with depth-first order.}
\label{fig:dfs-order}
\end{subfigure}\hfill
\begin{subfigure}[t]{0.49\textwidth}
\centering
\begin{tikzpicture}[x=1.2cm,y=0.7cm, gate/.style={circle,draw=orange!80!black,fill=orange!30,inner sep=1.7pt}]
  \coordinate (r) at (0,0);
  \coordinate (a) at (1.2,1.1);
  \coordinate (b) at (1.2,-1.1);
  \coordinate (t1) at (2.7,1.8);
  \coordinate (t2) at (2.7,0.5);
  \coordinate (t3) at (2.7,-0.5);
  \coordinate (t4) at (2.7,-1.8);

  \draw[thick] (r)--(a)--(t1);
  \draw[thick] (a)--(t2);
  \draw[thick] (r)--(b)--(t3);
  \draw[thick] (b)--(t4);

  \node[rootvertex,label=left:$r$] at (r) {};
  \node[steinervertex] at (a) {};
  \node[steinervertex] at (b) {};
  \node[terminal,label=right:$t_1$,inner sep=3pt] at (t1) {};
  \node[terminal,label=right:$t_2$,inner sep=1.4pt] at (t2) {};
  \node[terminal,label=right:$t_3$,inner sep=1.9pt] at (t3) {};
  \node[terminal,label=right:$t_4$,inner sep=2.4pt] at (t4) {};
  \node[gate] at (0,-0.25) {};
  \node[gate] at (2.6,1.55) {};
  \node[gate] at (2.8,1.55) {};
  \node[gate] at (2.6,-0.75) {};
  \node[gate] at (2.6,-2.05) {};
\end{tikzpicture}
\caption{The gates drawn at their anchors.}
\label{fig:gate-anchors}
\end{subfigure}
\medskip

\begin{subfigure}[t]{0.999\textwidth}
\centering
\begin{tikzpicture}[x=1.05cm,y=0.9cm,font=\small, gate/.style={circle,draw=orange!80!black,fill=orange!30,inner sep=1.7pt}]
  \fill[blue!10] (0,0) rectangle (3.7,0.62);
  \fill[green!10] (3.7,0) rectangle (5.0,0.62);
  \fill[orange!13] (5.0,0) rectangle (6.8,0.62);
  \fill[yellow!18] (6.8,0) rectangle (9.3,0.62);
  \fill[gray!18] (9.3,0) rectangle (11.0,0.62);
  \draw[thick] (0,0) rectangle (11.0,0.62);
  \draw[thick] (3.7,0)--(3.7,0.62) (5.0,0)--(5.0,0.62)
               (6.8,0)--(6.8,0.62) (9.3,0)--(9.3,0.62);

  \node at (1.85,0.31) {$w(t_1)$};
  \node at (4.35,0.31) {$w(t_2)$};
  \node at (5.90,0.31) {$w(t_3)$};
  \node at (8.05,0.31) {$w(t_4)$};
  \node[align=center,font=\scriptsize] at (10.15,0.31) {dummy weight\\at $r$};
  \node[below=2pt] at (0,0) {$0$};
  \node[below=2pt] at (9.3,0) {$W$};
  \node[below=2pt] at (11.0,0) {$mq$};

  \foreach \x in {0.7,2.9,5.1,7.3,9.5}{
    \draw[dashed,orange!65!black] (\x,0.62)--(\x,1.22);
    \node[gate] at (\x,0.92) {};
  }
  \node[orange!65!black,font=\scriptsize] at (0.70,1.5) {$\theta$};
  \node[orange!65!black,font=\scriptsize] at (2.90,1.50) {$\theta+q$};
  \node[orange!65!black,font=\scriptsize] at (5.10,1.5) {$\theta+2q$};
  \node[orange!65!black,font=\scriptsize] at (7.30,1.50) {$\theta+3q$};
  \node[orange!65!black,font=\scriptsize] at (9.50,1.5) {$\theta+4q$};

  \draw[-{Latex[length=2mm]}] (0,-0.62)--(11.0,-0.62);
\end{tikzpicture}
\caption{Corresponding accumulated weight for $(q,\theta)$.}
\label{fig:intervals}
\end{subfigure}
\caption{Gates on a four-terminal tree. We obtain $\Mark_q(\theta) = \{r, t_1, t_3, t_4\}$.}
\label{fig:interval-tree}
\end{figure}

We construct the gates as shown in \Cref{fig:interval-tree}: traverse the terminals of
$A_0$ in depth-first search order, keeping track of
their accumulated weight. The first gate
is anchored at the terminal where the accumulated
weight reaches $\theta$. Subsequently, a new gate
is anchored whenever the accumulated weight increases
by another $q$. We use
$m=\lceil W/q\rceil$ gates in total, where
$W=\sum_{t\in T}w(t)$. If the traversal ends before
all gates have been assigned, the remaining gates
are anchored at the root.

Let $N_{t,q}(\theta)$ denote the number of gates
anchored at terminal $t$. The modified weight function is
$w'(t):=qN_{t,q}(\theta)$.
Since the offset is chosen uniformly, each terminal
receives $w(t)/q$ gates in expectation. Hence,
$\E_\theta[w'(t)]=w(t)$.

We give a formal definition of the construction in
\Cref{sec:gate_pos}.

Later in \Cref{rem:derandomization} we explain how to obtain
the same approximation guarantee deterministically
by checking all relevant gate configurations.

\subsection{Motivation}

\textit{Why do we use gates?}
The gate construction is closely related to iterated tour partitioning for splittable capacitated vehicle routing~\cite{HaimovichRinnooyKan1985,AltinkemerGavish1987,BlauthTraubVygen2023}. The main difference is that we use a depth-first ordering of the terminals instead of a tour.

The connection to vehicle routing arises from interpreting the delay cost as a transportation cost. Each terminal's weight must be transported to the root along the final tree, and the delay cost is the length of that path multiplied by the amount transported on it.

Gates allow us to split this transportation into two stages. First, we transport the terminal weights to the gate anchors along the initial tree. We then transport the capacity of each gate to the root using the added shortcut. The sum of these two transportation costs provides an upper bound on the delay cost of the resulting tree.

\textit{Why do we need a direct and a pair candidate?}
The two candidates use different gate capacities, which give complementary bounds on the cost of transporting terminal weights to their gate anchors along the initial tree. At capacity $\mu$ in the direct candidate, a subtree incurs a large expected transportation cost when its total terminal weight is close to halfway between two consecutive multiples of $\mu$. But then this weight is close to a multiple of $\mu/2$, giving a small expected transportation cost for the pairwise candidate.

We can afford the smaller gate capacity $\mu/2$ for the pairwise candidate because pairwise reconnection is more efficient in terms of connection cost: two marked vertices can share a connection to the root. This allows us to mark roughly twice as many vertices without doubling the connection cost.

\textit{Where do we beat the decomposition gap?}
The analysis bounds the
pairwise matching cost by the cost of an edge-disjoint pairing in an
optimal Cost-Distance Steiner tree. This is valid because the minimum-cost matching cannot exceed the cost of any specific feasible matching. Thus, we can directly use the cost of an optimal Cost-Distance tree for that part of the analysis.

\section{Analysis: Gate networks}\label{sec:gates}

\subsection{Preliminaries}\label{sec:prelim}
The algorithm starts from a rooted tree $A_0$ with connection cost
\[
  C_0:=c(A_0),
  \qquad
  C_{\mathrm{ST}}\le C_0\le\omega C_{\mathrm{ST}}.
\]
In the spanning case $T=V\setminus\{r\}$, we take $A_0$ to be a minimum spanning tree, so
$\omega=1$ and $C_0=C_{\mathrm{ST}}$. For general terminal sets, Byrka et al.~\cite{ByrkaEtAl2013} and Traub and Zenklusen~\cite{TraubZenklusen2025} give deterministic
$\omega$-approximations for every fixed $\omega>\ln(4)$.

If $C_0=0$, the zero-length edges of $A_0$ connect every terminal to $r$, so $A_0$ is
optimal. If $D=0$, every positive-weight terminal has a zero-length edge to $r$ in the metric
closure. Adding these edges to $A_0$ and taking a shortest-path tree gives connection cost at
most $C_0$ and delay cost zero. Its approximation ratio is at most $\omega$, which is no larger
than our final bound. We therefore assume $C_0,D>0$ from now on.

\subsection{Definitions}\label{sec:gate_pos}
Fix $q>0$ and $\theta\in[0,q)$. Order the terminals as
$t_1,\ldots,t_n$ by a depth-first traversal of $A_0$ and set
\[
  p_0=0,\qquad p_i=\sum_{j=1}^i w(t_j),\qquad W=p_n,\qquad
  m=\left\lceil\frac{W}{q}\right\rceil.
\]
The terminal $t_i$ corresponds to
$I_i=[p_{i-1},p_i)$, and the dummy interval $I_{n+1}=[W,mq)$ is assigned to the root, as we saw in \Cref{fig:interval-tree}.

We use the set of anchored gates $G_q(\theta) = \{g_j: j = 0, \dots, m-1\}$ with anchor $a_{g_j}:=t_i$ if $\theta+jq\in I_i$ and $a_{g_j}:=r$ if $\theta+jq\in I_{n+1}$.

For $v\in T\cup\{r\}$, define the number of gates anchored at $v$ and its modified weight by
\[
  N_{v,q}(\theta):=\bigl|\{g\in G_q(\theta):a_g=v\}\bigr|,
  \qquad w'(v):=qN_{v,q}(\theta).
\]
The set of marked vertices is
\[
  \Mark_q(\theta):=
  \begin{cases}
    \{t\in T:N_{t,q}(\theta)>0\}, & \text{if $|\{t\in T:N_{t,q}(\theta)>0\}|$ is even},\\
    \{t\in T:N_{t,q}(\theta)>0\}\cup\{r\}, & \text{if $|\{t\in T:N_{t,q}(\theta)>0\}|$ is odd}.
  \end{cases}
\]
Thus the root is marked exactly when an odd number of non-root terminals are marked.

\subsection{Transporting weight to gate anchors}

We analyze the cost of transporting all terminal weight to the gate anchors while respecting the gate capacities.

We call a terminal $t$ \emph{below} an edge $e$ if removing $e$ from $A_0$ disconnects $t$ from the root. For every edge $e$ of $A_0$, the union of terminal weight intervals below e is connected and has length 
\[
  W_e:=\sum_{t\text{ below }e}w(t).
\]
Define the remainder
\[
  \rho_{e,q}:=W_e\bmod q,\qquad 0\le \rho_{e,q}<q.
\]
Let the number of gates anchored below $e$ be
\[
  N_{e,q}(\theta):=
  \bigl|\{g\in G_q(\theta):a_g\text{ is below }e\}\bigr|.
\]
We start by bounding the expected load imbalance between the terminal weight $W_e$ and the gate capacity $qN_{e,q}(\theta)$ below an edge $e$. This imbalance later determines how much weight has to be transported across $e$.

\begin{lemma}[Expected load imbalance]\label[lemma]{lem:residue}
For $\theta$ chosen uniformly from $[0,q)$,
\[
  N_{e,q}(\theta)\in
  \left\{
    \left\lfloor\frac{W_e}{q}\right\rfloor,
    \left\lceil\frac{W_e}{q}\right\rceil
  \right\}
 \qquad
    \text{and}
 \qquad
  \Pr\left[
    N_{e,q}(\theta)>
    \left\lfloor\frac{W_e}{q}\right\rfloor
  \right]
  =\frac{\rho_{e,q}}{q}.
\]
Moreover,
\[
  \E_{\theta}\!\left[|W_e-qN_{e,q}(\theta)|\right]
  =\frac{2\rho_{e,q}(q-\rho_{e,q})}{q}
  \le\frac q2.
\]
\end{lemma}

\begin{proof}
The terminal weight intervals below $e$ form a single interval of length $W_e$.
For $\theta$ in an interval of length $\rho_{e,q}$,
this interval contains $\lceil W_e/q\rceil$ gates;
for all other values of $\theta$, it contains
$\lfloor W_e/q\rfloor$ gates.

Consequently,
\[
  \begin{aligned}
  \E_{\theta}\left[|W_e-qN_{e,q}(\theta)|\right]
  &=\left(1-\frac{\rho_{e,q}}q\right)\rho_{e,q}
    +\frac{\rho_{e,q}}q(q-\rho_{e,q})\\
  &=\frac{2\rho_{e,q}(q-\rho_{e,q})}{q}
  \le\frac q2.
  \end{aligned}
\]
\end{proof}

We extend the weight function to the root by setting $w(r):=mq-W$.
\begin{definition}[Transportation plan]
A \emph{transportation plan} is a family
$f=(f_{t,g})_{t\in T\cup\{r\},\,g\in G_q(\theta)}$ of nonnegative values so that
\[
  \sum_{g\in G_q(\theta)}f_{t,g}=w(t)
  \quad(t\in T\cup\{r\}),
  \qquad
  \sum_{t\in T\cup\{r\}}f_{t,g}=q
  \quad(g\in G_q(\theta)).
\]
Here $f_{t,g}$ represents the weight transported from $t$ to the anchor $a_g$.
The optimal transportation cost over all transportation plans $f$ in $A_0$ is defined as
\[
  \Tr_q(\theta) := \min_{f}\sum_{t\in T\cup\{r\}}\sum_{g\in G_q(\theta)}
  f_{t,g}\dist_{A_0}(t,a_g).
\]
\end{definition}

\begin{lemma}[Transportation on a tree]\label[lemma]{lem:transport}
For every offset $\theta \in [0,q)$,
\begin{equation}\label{eq:transport-exact}
  \Tr_q(\theta)
  =\sum_{e\in E(A_0)}|W_e-qN_{e,q}(\theta)|c(e).
\end{equation}
In expectation,
\begin{equation}\label{eq:transport-expectation}
  \E_{\theta}[\Tr_q(\theta)]=\Phi_q, \qquad \Phi_q:=\sum_{e\in E(A_0)}
  \frac{2\rho_{e,q}(q-\rho_{e,q})}{q}c(e)
  \le\frac q2C_0.
\end{equation}
\end{lemma}
\begin{proof}
Deleting $e$ leaves an imbalance between terminal weight and gate capacity of $W_e-qN_{e,q}(\theta)$ in the component below it, so every feasible transportation plan must transport at least $|W_e-qN_{e,q}(\theta)|$ units across $e$. Conversely, orienting $A_0$ away from $r$ and assigning this signed flow to each edge satisfies every subtree balance; decomposing this flow into transportation paths attains the lower bound. This proves \eqref{eq:transport-exact}. Taking expectations and using \Cref{lem:residue} gives \eqref{eq:transport-expectation}.
\end{proof}

\subsection{From gate anchors to a feasible tree}
We combine the cost for transporting weight from terminals to gates with the cost for transporting the gate capacity to the root via a network of shortcuts $H$.
\begin{lemma}[Adding a connection network]\label[lemma]{lem:union}
Let $H$ be a graph connecting $r$ with every anchor $a_g$ for
$g\in G_q(\theta)$. A shortest-path tree $A\subseteq A_0\cup H$ satisfies
\begin{equation}\label{eq:union-bound}
  \cost(A)\le C_0+\Tr_q(\theta)+c(H)
  +q\sum_{g\in G_q(\theta)}\dist_H(r,a_g).
\end{equation}
\end{lemma}
\begin{proof}
A shortest-path tree of $A_0 \cup H$ has connection cost at most
$C_0+c(H)$. For an optimal transportation plan $f$,
\[
\begin{aligned}
  \sum_{t\in T}w(t)\dist_{A_0\cup H}(r,t)
  &\le
    \sum_{t\in T\cup\{r\}}\sum_{g\in G_q(\theta)}f_{t,g}
       \bigl(\dist_H(r,a_g)+\dist_{A_0}(a_g,t)\bigr)\\
  &=q\sum_{g\in G_q(\theta)}\dist_H(r,a_g)+\Tr_q(\theta).
\end{aligned}
\]
Combining the connection and delay bounds proves the claim.
\end{proof}

\section{Analysis: Direct and pairwise reconnection}\label{sec:reconnection}

\begin{lemma}[Expected anchor distance]\label[lemma]{lem:multiplicity}
For every rooted tree $F$ containing $T\cup\{r\}$,
\begin{equation}\label{eq:gate-distance}
  \E_{\theta}\!\left[
    \sum_{g\in G_q(\theta)}\dist_F(r,a_g)
  \right]
  =\frac1q\sum_{t\in T}w(t)\dist_F(r,t).
\end{equation}
\end{lemma}
\begin{proof}
Terminal $t$ anchors an expected $w(t)/q$ gates because its interval has length $w(t)$.
Root-anchored gates contribute zero. Multiplying the expected multiplicity at each terminal
by its root distance and summing proves the identity.
\end{proof}

\subsection{Direct reconnection}
For each $q>0$, let $\theta$ be uniformly distributed on $[0,q)$.
\begin{proposition}[Direct connection]\label[proposition]{prop:direct}
For every $q>0$, we have
\begin{equation}\label{eq:direct-candidate}
  \E_{\theta}[\cost(\Ad_q(\theta))]\le C_0+D+\frac Dq+\Phi_q.
\end{equation}
\end{proposition}
\begin{proof}
For a uniformly random offset $\theta$, let $H$ be the union of the direct metric edges
added by our algorithm. Thus
\[
  c(H) = \sum_{t\in \Mark_q(\theta)\setminus\{r\}}c(r,t) \le\sum_{g\in G_q(\theta)}c(r,a_g),
  \qquad
  \dist_H(r,a_g)=c(r,a_g).
\]
Since all anchors are connected with the root in $H$, applying \eqref{eq:union-bound} and taking expectations yields
\[
\begin{aligned}
  \E_{\theta}[\cost(\Ad_q(\theta))]
  &\le C_0+\E_{\theta}[\Tr_q(\theta)]
      +(1+q)\E_{\theta}\!\left[
        \sum_{g\in G_q(\theta)}c(r,a_g)
      \right]\\
  &\overset{\eqref{eq:transport-expectation},\eqref{eq:gate-distance}}{=}
    C_0+\Phi_q+(1+q)\frac Dq.
\end{aligned}
\]
\end{proof}
By \Cref{lem:transport}, we have $\Phi_q \le qC_0/2$, so the proposition shows that $\Ad_q(\theta)$ matches the Foos--Held--Spitzley guarantee~\cite{FoosHeldSpitzley2023Conference}.

\subsection{Pairwise reconnection}\label{sec:pairing_gates}

For $q>0$, let $\theta$ be uniformly distributed on $[0,q)$.
The pairwise candidate $\Ap_q(\theta)$ uses a perfect matching on the vertices in $\Mark_q(\theta)$.

Recall that $w'(v)$ is defined as $q$ times the number of gates anchored at $v$.
For distinct $u,v\in\Mark_q(\theta)$, the matching edge cost is
\begin{equation}\label{eq:marked-pair-cost}
  K_{u,v}=
  \min_F\{c(F)+w'(u)\dist_F(r,u)+w'(v)\dist_F(r,v)\},
\end{equation}
where $F$ ranges over rooted trees spanning $r,u,v$. Write $F_{u,v}$ for a
minimizing tree.

\begin{lemma}[Cost of a two-terminal connection]\label[lemma]{lem:two-terminal}
For distinct $u,v\in\Mark_q(\theta)$, the matching edge cost $K_{u,v}$ in \eqref{eq:marked-pair-cost} and a realizing tree can
be found by enumerating the $|V|$ possible branch vertices. If $r,u,v$ are distinct, then
\begin{equation}\label{eq:weighted-pair-cost}
\begin{aligned}
  K_{u,v}=\min_{z\in V}\Bigl\{
    &(1+w'(u)+w'(v))c(r,z)\\
    &+(1+w'(u))c(z,u)+(1+w'(v))c(z,v)\Bigr\}.
\end{aligned}
\end{equation}
\end{lemma}
A realizing tree uses at most three edges when $r,u,v$ are distinct, and one edge when one marked vertex is $r$.
\begin{proof}
For pairwise distinct $r,u,v$, any minimal subtree consists of three paths meeting at one vertex. Replacing each path by its metric edge cannot
increase the objective.
If $u$ or $v$ equals $r$, the optimum is the direct connection to the other
terminal.
\end{proof}

\begin{figure}[htbp]
\centering
\begin{tikzpicture}[x=1.20cm,y=1.00cm]
  \coordinate (r) at (0,0);
  \coordinate (a) at (0,1);
  \coordinate (t1) at (-1,1.5);
  \coordinate (b) at (1,1);
  \coordinate (t2) at (2,1);
  \coordinate (c) at (1,2);
  \coordinate (t3) at (0,2);
  \coordinate (d) at (1,3);
  \coordinate (t4) at (0.5,3.5);
  \coordinate (t5) at (1.5,3.5);
  \draw[thick,gray!60] (a)--(b) (c)--(d);
  \draw[pairingpath] (r)--(a)--(t1);
  \draw[pairingpath] (t2)--(b)--(c)--(t3);
  \draw[pairingpath] (t4)--(d)--(t5);
  \node[rootvertex,markedvertex,label=below:$r$] at (r) {};
  \node[steinervertex] at (a) {};
  \node[steinervertex] at (b) {};
  \node[steinervertex] at (c) {};
  \node[steinervertex] at (d) {};
  \node[terminal,markedvertex,label=left:$t_1$] at (t1) {};
  \node[terminal,markedvertex,label=right:$t_2$] at (t2) {};
  \node[terminal,markedvertex,label=left:$t_3$] at (t3) {};
  \node[terminal,markedvertex,label=above left:$t_4$] at (t4) {};
  \node[terminal,markedvertex,label=above right:$t_5$] at (t5) {};
\end{tikzpicture}
\caption{An edge-disjoint pairing of marked vertices in the optimal tree $A^*$.
Orange paths realize the pairs $\{r,t_1\}$, $\{t_2,t_3\}$, and
$\{t_4,t_5\}$.}
\label{fig:edge-disjoint-pairing}
\end{figure}
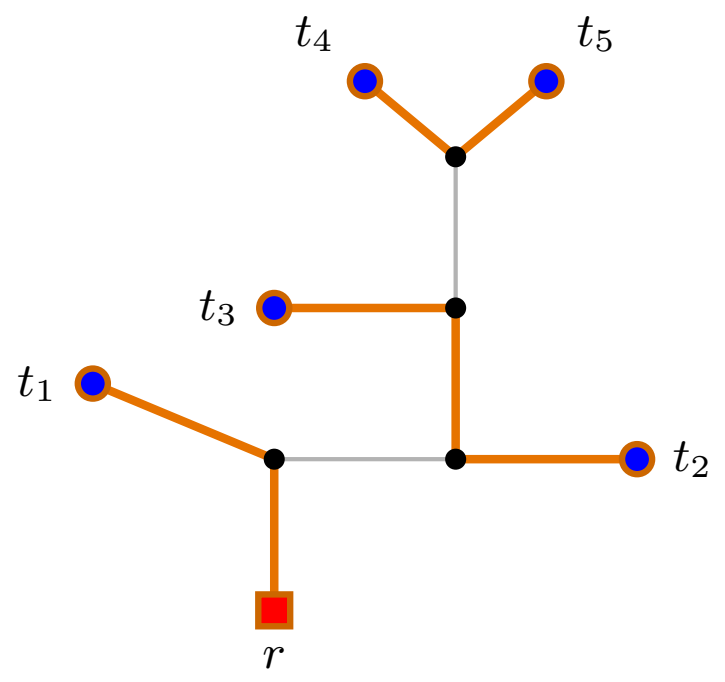

We bound the cost of the matching by the cost of a
feasible pairing of the marked vertices that exists inside an optimal tree using
edge-disjoint paths, similarly to the matching-based analysis
of Meyerson, Munagala, and Plotkin~\cite{MeyersonMunagalaPlotkin2008}.
An example is shown in \Cref{fig:edge-disjoint-pairing}.

\begin{lemma}[Pairing marked vertices in a tree]\label[lemma]{lem:tree-pairing}
Given an even set of marked vertices in a rooted tree $A$, they can be
partitioned into pairs $P$ such that the $u$--$v$ paths in $A$, for
$\{u,v\}\in P$, are edge-disjoint.
\end{lemma}
\begin{proof}
Transform \(A\) into a binary tree \(A'\) by attaching a new marked leaf to each marked vertex, unmarking the original vertices, and splitting vertices of high degree using auxiliary edges.

Let $E'$ contain exactly those edges with an odd number of marked vertices below them. Then every internal vertex has even degree in $E'$, while every marked leaf has degree one. Thus $E'$ is a disjoint union of paths with marked endpoints.

Contracting the auxiliary edges gives the desired edge-disjoint paths in $A$.
\end{proof}

\begin{proposition}[Pair matching]\label[proposition]{prop:matching}
For every $q>0$, we have
\begin{equation}\label{eq:matching-candidate}
  \E_{\theta}[\cost(\Ap_q(\theta))]\le
  C_0+\Phi_q+\frac12C^*+\left(1+\frac1{2q}\right)D^*.
\end{equation}
\end{proposition}
\begin{proof}
Fix an offset $\theta$ and apply \Cref{lem:tree-pairing} to $\Mark_q(\theta)$ in $A^*$. For the
resulting pairing $P$,
\[
  \sum_{\{u,v\}\in P}\dist_{A^*}(u,v)\le C^*,
\]
because the corresponding paths are edge-disjoint. For a pair
$\{u,v\}\in P$, let $L_{u,v}$ be the minimal subtree of $A^*$ containing
$r,u,v$. Then
\[
\begin{aligned}
  K_{u,v}
  &\le c(L_{u,v})
      +w'(u)\dist_{A^*}(r,u)+w'(v)\dist_{A^*}(r,v)\\
  &=\frac12 \dist_{A^*}(u,v) +
      \Bigl(\frac{1}{2} +w'(u)\Bigr)\dist_{A^*}(r,u)+\Bigl(\frac{1}{2} +w'(v)\Bigr)\dist_{A^*}(r,v).
\end{aligned}
\]
The cost of the pairing is at most
\[
\begin{aligned}
  \sum_{\{u,v\}\in P}K_{u,v}
  &\le \frac12C^* + \sum_{\{u,v\}\in P}\sum_{z\in\{u,v\}}
      \Bigl(\frac12+w'(z)\Bigr)\dist_{A^*}(r,z) \\
  & = \frac12C^*+\sum_{v \in \Mark_q(\theta)}\Bigl(\frac{1}{2} +qN_{v,q}(\theta)\Bigr)\dist_{A^*}(r,v) \\
  & \le \frac12C^*+\Bigl(\frac{1}{2} +q\Bigr)\sum_{g \in G_q(\theta)} \dist_{A^*}(r,a_g).
\end{aligned}
\]
Let $M$ be the minimum-cost perfect matching on
$\Mark_q(\theta)$, and let $H$ be the union of its realizing trees $F_{u,v}$.
By the definition of $w'$, and because taking the union can only reduce the
sum of connection costs and the root distances,
\[
  c(H)+q\sum_{g\in G_q(\theta)}\dist_H(r,a_g)
   \le \sum_{\{u,v\}\in M}K_{u,v}
  \le\sum_{\{u,v\}\in P}K_{u,v}.
\]
Combining this with the previous inequality, inserting the resulting bound into \eqref{eq:union-bound}, and taking expectations gives
\[
\begin{aligned}
  \E_{\theta}[\cost(\Ap_q(\theta))]
  &\le C_0+\Phi_q+\frac12C^*
     +\left(q+\frac12\right)
       \E_{\theta}\!\left[
         \sum_{g\in G_q(\theta)}\dist_{A^*}(r,a_g)
       \right]\\
  &\overset{\eqref{eq:gate-distance}}{=}
    C_0+\Phi_q+\frac12C^*+\left(1+\frac1{2q}\right)D^*.
\end{aligned}
\]
\end{proof}

\Cref{alg:shifted-gates} computes a value $\mu \in \R$ and uses the values $\mu$ and $\mu/2$ for its candidates.
The remaining proof analyzes the relation between $\Phi_\mu$ and $\Phi_{\mu/2}$.

\section{Analysis: Final approximation guarantees}

\subsection{Complementary gate capacities}
The key idea is that, for a given $\mu>0$, a subtree has large expected transportation cost at capacity $\mu$ only when its
total terminal weight is near an odd multiple of $\mu/2$. Its weight is then near a
multiple of $\mu/2$, where the expected transportation cost at half-capacity is small:

For $W\ge0$, define
\[
  \tau_\mu(W)=\frac{2\rho_\mu(\mu-\rho_\mu)}\mu,
  \qquad
  \rho_\mu=W\bmod \mu.
\]
Thus $\Phi_\mu=\sum_e\tau_\mu(W_e)c(e)$. For every $W$, we have
$\tau_\mu(W)\le\mu/2$, with equality when $\rho_\mu=\mu/2$. Let
\[
  S_e:=\frac\mu2-\tau_\mu(W_e),
  \qquad
  S:=\sum_e S_e c(e).
\]
The quantity $S$ is the total saving relative to the worst-case potential for $\mu$.

\begin{lemma}[The relation between $\mu$ and $\mu/2$]\label[lemma]{lem:two-scale}
For every edge $e$,
\[
  \tau_{\mu/2}(W_e)
  =2\sqrt{\frac\mu2S_e}-2S_e.
\]
Consequently,
\begin{align}
  \Phi_\mu&=\frac\mu2C_0-S,\label{eq:Phi-mu}\\
  \Phi_{\mu/2}&\le
  2\sqrt{\frac\mu2SC_0}-2S.\label{eq:Phi-half}
\end{align}
\end{lemma}

\begin{proof}
Summing $\tau_\mu(W_e)=\mu/2-S_e$ over all edges gives \eqref{eq:Phi-mu}.

Put $y=W_e\bmod\mu$ and $d=|y-\mu/2|$. Then
\[
  S_e=\frac{\mu}{2}-\frac{2y(\mu-y)}{\mu}
  =\frac{2\left(y-\mu/2\right)^2}{\mu},
  \qquad d=\sqrt{\frac\mu2S_e}.
\]
Upon reducing $y$ modulo $\mu/2$, the resulting residue is either $d$ or
$\mu/2-d$. Therefore,
\[
  \tau_{\mu/2}(W_e)=\frac{2d(\mu/2-d)}{\mu/2}
  =2\sqrt{\frac\mu2S_e}-2S_e.
\]
Hence,
\[
\begin{aligned}
\Phi_{\mu/2}
  &= 2\sqrt{\frac\mu2}\sum_e \sqrt{S_e}\,c(e) - 2S \\
  &\le
    2\sqrt{\frac\mu2}\,
    \sqrt{\biggl(\sum_e S_e c(e)\biggr)
          \biggl(\sum_e c(e)\biggr)}
    - 2S
    && \text{(Cauchy--Schwarz)} \\
  &= 2\sqrt{\frac\mu2SC_0} - 2S.
\end{aligned}
\]
This is \eqref{eq:Phi-half}.
\end{proof}

\subsection{Combining the guarantees}\label{sec:steiner-extension}

Fix $\lambda>0$ and set
\[
  \mu:=\lambda\sqrt{\frac{2D}{C_0}}.
\]
Let $\theta$ and $\theta'$ be the offsets sampled in \Cref{alg:shifted-gates}.
The two candidates are $\Ad_{\mu}(\theta)$ and $\Ap_{\mu/2}(\theta')$.
Let $S$ be the saving from
\Cref{lem:two-scale} at this capacity. Substituting
\eqref{eq:Phi-mu} and \eqref{eq:Phi-half} into
\eqref{eq:direct-candidate} and \eqref{eq:matching-candidate} gives
\begin{align}
  \E[\cost(\Ad_{\mu}(\theta))]
  & \le C_0+D+\frac{D}{\mu}+\frac{\mu C_0}{2}-S
  \notag\\
  & = C_0+D+\left(\lambda+\frac1\lambda\right)\sqrt{\frac{DC_0}{2}}-S, \label{eq:unnormalized-direct}\\
  \E[\cost(\Ap_{\mu/2}(\theta'))]
  & \le C_0+2\sqrt{\frac{\mu}2SC_0}-2S
     +\frac12C^*+\left(1+\frac1{\mu}\right)D^* \notag\\
   & =C_0+2\sqrt{\lambda S\sqrt{\frac{DC_0}{2}}}-2S
    +\frac12C^*+D^*+\frac{D^*}{\lambda}\sqrt{\frac{C_0}{2D}}.
  \label{eq:unnormalized-matching}
\end{align}

Set
\[
  L:=C_{\mathrm{ST}}+D,
  \qquad
  x:=\frac{C_{\mathrm{ST}}}{L},
  \qquad
  p:=\frac{\OPT}{L}\ge1,
  \qquad
  u:=\frac{2S}{\mu\,\omega C_{\mathrm{ST}}}\in[0,1],
\]
and abbreviate
\[
  g_\omega(x):=\sqrt{2\omega x(1-x)},
  \qquad
  h_{\omega,\lambda}(x)
  :=\frac1\lambda\sqrt{\frac{\omega x}{2(1-x)}}
  =\frac1{\mu}.
\]

\begin{lemma}[Approximation bounds]\label[lemma]{lem:steiner-normalized}
Define
\begin{align*}
  A_{\omega,\lambda}(x,u)
  &:=1+(\omega-1)x
    +g_\omega(x)\left(\frac1{2\lambda}
      +\frac\lambda2(1-u)\right),
  \\
  B_{\omega,\lambda}(x,u)
  &:=1+\left(\omega-\frac12\right)x
    +g_\omega(x)\left(\frac1{2\lambda}
      +\lambda(\sqrt u-u)\right).
\end{align*}
Then
\[
  \frac{\E[\cost(\Ad_{\mu}(\theta))]}{\OPT}
  \le\frac{A_{\omega,\lambda}(x,u)}p
\]
and
\[
  \frac{\E[\cost(\Ap_{\mu/2}(\theta'))]}{\OPT}
  \le 1+h_{\omega,\lambda}(x)
  +\frac{B_{\omega,\lambda}(x,u)-1-h_{\omega,\lambda}(x)}p.
\]
\end{lemma}

\begin{proof}
For fixed $D,S,C^*,D^*$, the right-hand sides of both \eqref{eq:unnormalized-direct} and \eqref{eq:unnormalized-matching} are increasing in $C_0$.
Increasing $C_0$ preserves the valid range of $S$ because
\[
  0\le S\le\frac{\mu C_0}{2}
  =\frac{\lambda}{2}\sqrt{2DC_0},
\]
where the right side increases with $C_0$.
Therefore, it suffices to evaluate
\eqref{eq:unnormalized-direct} and \eqref{eq:unnormalized-matching} at
\(C_0=\omega C_{\mathrm{ST}}\).

Using \(D=(1-x)L\),
\(\omega C_{\mathrm{ST}}=\omega xL\), and
\[
  \sqrt{\frac{D\omega C_{\mathrm{ST}}}{2}}
  =\frac{g_\omega(x)L}{2},
  \qquad
  S=\frac{u\lambda g_\omega(x)L}{2},
\]
we obtain
\begin{align*}
  \frac{\E[\cost(\Ad_{\mu}(\theta))]}{L}
  &\le \omega x
  +(1-x)
  +\frac{\lambda+\lambda^{-1}}{2}g_\omega(x)
  -\frac{u\lambda}{2}g_\omega(x) =A_{\omega,\lambda}(x,u).
\end{align*}
Dividing by $p$ and using \(\OPT=pL\) gives the first bound.

For the second bound, subject to \(C^*+D^*=pL\) and
\(C^*\ge C_{\mathrm{ST}}\), the right-hand side is maximized when
\(C^*=C_{\mathrm{ST}}=xL\) and \(D^*=(p-x)L\).
\[
  2\sqrt{\frac{\mu}{2}S\omega C_{\mathrm{ST}}}-2S
  =\lambda g_\omega(x)L(\sqrt{u}-u), \qquad
  \frac{g_\omega(x)}{2\lambda}=(1-x)h_{\omega,\lambda}(x).
\]
With \(h_{\omega,\lambda}(x)=1/\mu\), substitution gives
\begin{align*}
  \frac{\E[\cost(\Ap_{\mu/2}(\theta'))]}{L}
  &\le
  \omega x+\lambda g_\omega(x)(\sqrt{u}-u)
  +\frac{x}{2}+(1+h_{\omega,\lambda}(x))(p-x) \\
  &=(1+h_{\omega,\lambda}(x))p
  +\left(\omega-\frac12-h_{\omega,\lambda}(x)\right)x
  +\lambda g_\omega(x)(\sqrt{u}-u) \\
  &=(1+h_{\omega,\lambda}(x))p+B_{\omega,\lambda}(x,u)-1-h_{\omega,\lambda}(x).
\end{align*}
Dividing by $p$ yields the second bound.
\end{proof}

\begin{theorem}[Parameterized approximation guarantee]\label{thm:general-factor}
Let $\omega\ge1$, and suppose that a polynomial-time
$\omega$-approximation for Steiner Tree is available. For every fixed
positive rational $\lambda$, \Cref{alg:shifted-gates} can be implemented in polynomial time and returns a random feasible
tree $A$ satisfying
\[
  \E[\cost(A)]\le\mathcal R_{\omega,\lambda}\,\OPT,
\]
where
\begin{equation}\label{eq:general-factor-definition}
\begin{aligned}
  \mathcal R_{\omega,\lambda}:=
  \sup_{\substack{0<x<1,\ 0\le u\le1,\ p\ge1}}
  \min\Biggl\{
  &\frac{A_{\omega,\lambda}(x,u)}p,\\
  &1+h_{\omega,\lambda}(x)
   +\frac{B_{\omega,\lambda}(x,u)-1-h_{\omega,\lambda}(x)}p
  \Biggr\}.
\end{aligned}
\end{equation}
\end{theorem}

\begin{proof}
The algorithm returns the cheaper of the two random candidates. Therefore
\[
\begin{aligned}
  \E[\cost(A)]
  &\le\min\{\E[\cost(\Ad_{\mu}(\theta))],\E[\cost(\Ap_{\mu/2}(\theta'))]\}.
\end{aligned}
\]
Applying \Cref{lem:steiner-normalized} proves the approximation guarantee.

The assumed Steiner-tree approximation can be computed in polynomial time. In the spanning case,
it is a minimum spanning tree. Each pair cost $K_{u,v}$ is found by $O(|V|)$ enumerations, and a minimum-cost perfect matching
can be computed in polynomial time.
We add at most one direct connection or one pair tree for every terminal. A shortest-path tree of the union is computed in
polynomial time.

We do not explicitly compute anchors for all gates but only the number of gates at each vertex.
For rational input data of polynomial encoding length, $\mu$ may be irrational. Nevertheless,
all quantities used by the algorithm can be represented as $a+b\mu$ with
$a,b\in\mathbb Q$, and this representation is closed under the required arithmetic operations.
Comparisons of quantities of the form $a+b\mu$ can be performed exactly in polynomial time, since $\mu^2\in\mathbb Q$. As the $O(n^3)$ weighted matching algorithm~\cite{KorteVygen2018} uses only arithmetic operations and comparisons, it can also be implemented with polynomial bit complexity.
\end{proof}

\begin{remark}[Derandomization]\label{rem:derandomization}
For $q\in\{\mu,\mu/2\}$ and the interval boundaries $p_i$ from \Cref{sec:gate_pos}, put
\[
  r_i(q):=p_i-q\left\lfloor\frac{p_i}{q}\right\rfloor,
  \qquad
  \Theta_q:=\{r_i(q):0\le i\le n\}.
\]
For each capacity $q\in\{\mu,\mu/2\}$, evaluate its corresponding candidate
at all offsets in $\Theta_q$ and retain the cheapest tree overall. There are
only $O(|T|)$ offsets per capacity, and the best candidate at a capacity
costs no more than the expected value. Hence this gives a
deterministic polynomial-time algorithm with the same approximation
guarantee.
\end{remark}

For the spanning-case guarantee stated in \Cref{sec:contributions}, set $\omega=1$ and
$\lambda=11/10$ in \Cref{thm:general-factor}. The calculation in
\Cref{sec:one-dimensional-certification} gives
$\mathcal R_{1,11/10}<1.698$.

For the general-terminal guarantee, note that
$\ln(4)<1.3863$, so the result of Byrka et al.~\cite{ByrkaEtAl2013} provides a suitable Steiner tree approximation. We then apply \Cref{thm:general-factor} with $\omega=1.3863$ and $\lambda=43/40$. The calculation in
\Cref{sec:one-dimensional-certification} gives
$\mathcal R_{1.3863,43/40}<2.041$.

\section{Lower bounds for the reconnection algorithms}\label{sec:reconnection-lower-bound}

\begin{theorem}[Reconnection lower bounds]\label{thm:reconnection-lb}
For every $\varepsilon>0$, there are spanning instances in the $\ell_1$ metric such that
\begin{enumerate}[label=(\roman*),leftmargin=*]
  \item cut-and-reconnect algorithms that start with a minimum spanning tree have an approximation ratio of at least
        $1+1/\sqrt{2}-\varepsilon$; \label{item:i}
  \item the algorithm presented in this paper has a worst-case approximation ratio of at least
        $3/2-\varepsilon$.
\end{enumerate}
\end{theorem}

\begin{proof}
For $a\in\{\sqrt{2},2\}$ and an integer $s\ge1$, construct the staircase instance in
\Cref{fig:hard-staircase} with $s$ steps. For $i=0,\ldots,s-1$, column $i$ has horizontal coordinate $ai$ and contains vertices at
all integer heights from $0$ to $i+1$. The last column has horizontal coordinate $as$ and
contains vertices at all integer heights from $0$ to $s$. All vertices are terminals, except $(0,0)$, which is the root. For each
$i=0,\ldots,s-1$, place a terminal at $(a(i+1/2),i+1)$. We consider the complete graph on these vertices, with every edge length given by the
$\ell_1$ distance between its endpoints. The bottom terminals $t_i$ at $(ai,0)$, for $i=1,\ldots,s$, have
weight $1/a$, and all other terminals have weight zero.

Define the tree $A_s$ as follows. For each $i=1,\ldots,s$, it contains the two horizontal
segments joining $(a(i-1),i)$ to $(ai,i)$ through the midpoint terminal, as well as the $i$ unit
vertical segments from $(a(i-1),0)$ to $(a(i-1),i)$. It also contains the $s$ unit vertical
segments from $(as,0)$ to $(as,s)$.

\begin{figure}[htbp]
\centering
\begin{subfigure}[t]{0.48\textwidth}
\centering
\begin{tikzpicture}[xscale=0.86,yscale=0.70]
  \foreach \i in {1,2,3,4,5} {
    \draw (\i-1,0)--(\i-1,\i)--(\i,\i);
    \node[terminal] at (\i,0) {};
    \node[smallterminal] at (\i,\i) {};
    \node[smallterminal] at (\i-1,\i) {};
    \node[smallterminal] at (\i-0.5,\i) {};
    \foreach \j in {1,...,\i} {
      \node[smallterminal] at (\i,\j) {};
    }
  }
  \draw (5,0)--(5,5);
  \path (0,0)--node[below] {$a$} ++(1,0);
  \path (0,0)--node[left] {$1$} ++(0,1);
  \node[rootvertex] at (0,0) {};
  \node[terminal,label=below:{\color{blue}$1/a$}] at (5,0) {};
  \node[smallterminal,label=right:{\color{blue}$0$}] at (5,1) {};
\end{tikzpicture}
\caption{The initial minimum spanning tree $A_s$.}
\label{fig:hard-mst}
\end{subfigure}\hfill
\begin{subfigure}[t]{0.48\textwidth}
\centering
\begin{tikzpicture}[xscale=0.86,yscale=0.70]
  \draw (0,0)--(5,0);
  \draw (0,0)--(0,1);
  \foreach \i in {1,2,3,4,5} {
    \draw (\i-1,1)--(\i-1,\i)--(\i,\i);
    \node[terminal] at (\i,0) {};
    \node[smallterminal] at (\i,\i) {};
    \node[smallterminal] at (\i-1,\i) {};
    \node[smallterminal] at (\i-0.5,\i) {};
    \foreach \j in {1,...,\i} {
      \node[smallterminal] at (\i,\j) {};
    }
  }
  \draw (5,1)--(5,5);
  \path (0,0)--node[below] {$a$} ++(1,0);
  \path (0,0)--node[left] {$1$} ++(0,1);
  \node[rootvertex] at (0,0) {};
  \node[terminal,label=below:{\color{blue}$1/a$}] at (5,0) {};
  \node[smallterminal,label=right:{\color{blue}$0$}] at (5,1) {};
\end{tikzpicture}
\caption{A comparison tree of cost $s^2+O_a(s)$.}
\label{fig:hard-opt}
\end{subfigure}
\caption{Staircase instance with $s=5$ steps; only bottom vertices $t_i=(ai,0)$ have positive weight.}
\label{fig:hard-staircase}
\end{figure}
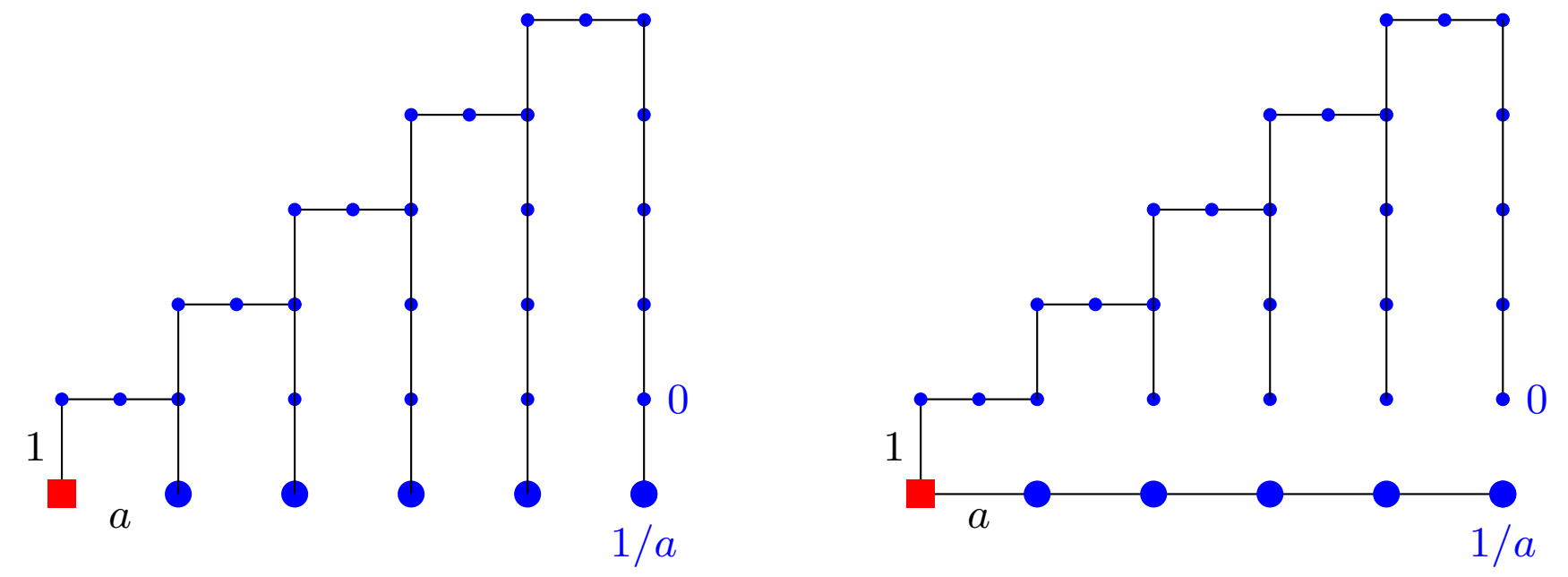

\begin{lemma}\label[lemma]{lem:hard-mst}
For $a\in\{\sqrt{2},2\}$, the tree $A_s$ is the unique minimum spanning tree
on the displayed vertices.
\end{lemma}
\begin{proof}
All edges in $A_s$ have length at most $1$, whereas every edge not contained in $A_s$ has
length greater than $1$. Hence each edge of $A_s$ is the unique minimum-length edge across
its fundamental cut.
\end{proof}

The connection cost of $A_s$ is
\[
  c(A_s)=as+\frac{s(s+1)}2+s.
\]
The path from $r$ to $t_i$ in $A_s$ has length $(a+2)i$, and therefore
\[
  \sum_{i=1}^s\frac1a\dist_{A_s}(r,t_i)
  =\left(1+\frac2a\right)\frac{s(s+1)}2.
\]
Consequently,
\[
  \cost(A_s)=\left(1+\frac1a\right)s^2+O_a(s).
\]

\begin{lemma}[Geometry of a shortcut]\label[lemma]{lem:hard-geometry}
A root-connected network $H$ added to $A_s$ with $k$ non-root vertices can shorten the distances
to at most $k$ bottom terminals. Any root-connected network that shortens the distance to $t_i$ has connection cost
at least $ai$. The decrease in root distance for each such terminal $t_i$ is
at most $2i$.
\end{lemma}
\begin{proof}
For each $i$, the vertices $(ai,h)$ with $h<i$ are connected to
the rest of $A_s$ only through $(ai,i)$. The path from $r$ to
$(ai,i)$ in $A_s$ is an $\ell_1$ shortest path. Thus, shortening
the distance to $t_i$ requires $H$ to contain a vertex $(ai,h)$
with $h<i$. Since these vertices are distinct for different $i$,
at most $k$ bottom terminals can benefit.

Such a vertex has distance at least $ai$ from $r$, so any
root-connected network helping $t_i$ has cost at least $ai$.
Finally, $\dist_{A_s}(r,t_i)=(a+2)i$ and $c(r,t_i)=ai$,
so the distance decrease is at most $2i$.
\end{proof}

\textbf{Cut-and-reconnect algorithms.}
Set $a=\sqrt{2}$. The algorithm deletes edges of $A_s$ and reconnects every
resulting non-root component by a single edge from the root. Let $K$ be a non-root component containing no positive-weight terminal, and
let $f$ be the unique deleted edge of $A_s$ joining $K$ to the component
closer to the root. Let $e$ be the edge used to reconnect $K$ to the root.
Since $e$ crosses the fundamental cut of $f$, \Cref{lem:hard-mst} gives
$c(f)\le c(e)$. Replacing $e$ by $f$ merges $K$ with its parent component. This does not increase the connection cost and does not
change any positive-weight terminal's root distance. Repeating the
replacement eliminates every component without a positive-weight terminal.
Thus every non-root component contains one of $t_1,\ldots,t_s$, and we may
assume that there are at most $s+1$ components.

Each non-root component has one new root edge. By
\Cref{lem:hard-geometry}, each edge $e$ can shorten the distance of at most
one bottom terminal $t_j$. The lemma bounds the delay saving
by
\[
  \frac1a\,2j=aj\le c(e).
\]
Thus the reconnections themselves produce no net saving.
At most $s$ original edges are deleted, each of constant length. Denote the returned tree by
$A_{\mathrm{CR}}$. It therefore has cost
\begin{equation}\label{eq:foos-lb}
  \cost(A_{\mathrm{CR}})
  \ge \cost(A_s) - O(s) = \left(1+\frac1{\sqrt2}\right)s^2-O(s).
\end{equation}

\textbf{The algorithm of this paper.}
Set $a=2$. Zero-weight terminals never anchor a gate, so every marked non-root
vertex is one of the bottom terminals $t_1,\ldots,t_s$. The following bounds hold
for every capacity and offset.

For the first candidate $\Ad$, a direct edge from $r$ to an anchor $t_i$ has connection cost $2i$ and, by
\Cref{lem:hard-geometry}, decreases the delay by at most $i$. Thus the direct shortcuts do not
create a saving.

For $\Ap$, consider a matched pair $t_i,t_j$ with $i<j$. Every rooted tree spanning
$r,t_i,t_j$ has connection cost at least $2j$ and auxiliary delay
at least $2iw'(t_i)+2jw'(t_j)$. The path with edges $\{r,t_i\}$ and
$\{t_i,t_j\}$ attains both lower bounds. In the branch-vertex formula
\eqref{eq:weighted-pair-cost}, we set
$z=t_i$, so this is the tree $F_{t_i,t_j}$ selected by the algorithm.
A pair consisting of $r$ and $t_j$ is realized by the direct edge $\{r,t_j\}$.

Because the matching is on distinct marked vertices, the realizing trees of different
matched pairs have no common edge. By
\Cref{lem:hard-geometry}, the tree for anchors $t_i,t_j$ can reduce the delay only for those two
bottom terminals. Its connection cost is $2j$, whereas its delay saving is at most
$i+j\le2j$. The same holds for an $r$--$t_j$ pair by taking $i=0$. Summing over
the matching, the pairwise augmentation does not improve the objective before edges of $A_s$
are removed.

There are at most $s+1$ marked vertices, so either candidate adds only $O(s)$ edges to $A_s$. Before a
shortest-path tree is extracted, the augmentation yields no net
reduction in the objective. A shortest-path tree preserves all
root distances in the union. Adding $k$ edges to a tree permits a spanning tree of
the union to omit at most $k$ original edges. Every edge of $A_s$ has
constant length, so extracting the shortest-path tree reduces the connection cost by
only $O(s)$. Hence,
\begin{equation}\label{eq:our-lb}
  \min\{\cost(\Ad_{\mu}(\theta)),\cost(\Ap_{\mu/2}(\theta'))\}
  \ge \cost(A_s)-O(s)=\frac32s^2-O(s).
\end{equation}

\textbf{Comparison with the optimum.}
The tree in \Cref{fig:hard-opt} contains all bottom segments of combined length $as$ and a
shifted copy of the staircase. Its connection cost is $2as+s(s+1)/2$, and every $t_i$ has root distance $ai$.
Hence
\begin{equation}\label{eq:hard-opt-ub}
  \OPT\le 2as+\frac{s(s+1)}2+\frac{s(s+1)}2=s^2+O_a(s).
\end{equation}
Combining \eqref{eq:foos-lb} with \eqref{eq:hard-opt-ub} for $a=\sqrt{2}$ gives a ratio tending
to $1+1/\sqrt{2}$. Combining \eqref{eq:our-lb} with \eqref{eq:hard-opt-ub} for $a=2$ gives a
ratio tending to $3/2$. Choosing $s$ sufficiently large proves both statements.
\end{proof}

\section{Relation to the Cable-Trench Problem}\label{app:cable-trench}

In the \textsc{Cable-Trench Problem}~\cite{VaskoEtAl2002}, a set of computers, represented by
terminals, must be connected to a server, represented by the root. Laying a cable requires a
trench, but several cables may use the same trench. Cable and trench costs are usually assumed
to be proportional. In the Steiner-tree version, trenches may merge at non-terminal
vertices~\cite{BeneditoEtAl2023}. Other variants use separate cost functions for cables and
trenches~\cite{VaskoEtAl2016}.

For parameters $\lambda_1,\lambda_2\ge0$, the objective of the Cable-Trench Problem is
\[
  \operatorname{ct}(A)=\lambda_1 c(A)+\lambda_2\sum_{t\in T}\dist_A(r,t).
\]
When $\lambda_1>0$, this is a special case of the Uniform Cost-Distance Steiner Tree Problem: divide the objective by
$\lambda_1$ and set $w(t)=\lambda_2/\lambda_1$ for every terminal. The converse is the
content of \Cref{thm:cable-reduction}.

\begin{theorem}[Reduction to Cable-Trench with Steiner points]\label{thm:cable-reduction}
Suppose that the Steiner tree version of the Cable-Trench Problem has an $\alpha$-approximation. Then, for
every fixed $\varepsilon>0$, the Uniform Cost-Distance Steiner Tree Problem has an
$(\alpha+\varepsilon)$-approximation. For fixed $\alpha$, the running time is polynomial in the input size and in
$1/\varepsilon$.
\end{theorem}

\begin{proof}
Consider an instance of the Uniform Cost-Distance Steiner Tree Problem. We may assume $0<\varepsilon\le1$. Call a
terminal heavy if its weight is at least $1/\varepsilon$, and temporarily remove all heavy
terminals from the terminal set, retaining their vertices as possible Steiner vertices. For every remaining terminal, round its weight up to the next multiple of
$\varepsilon/(\alpha|T|)$. Replace a zero weight by $\varepsilon/(\alpha|T|)$. Replace
a terminal of rounded weight $k\varepsilon/(\alpha|T|)$ by $k$ copies connected by zero-length edges, and then take the metric closure. The new
terminal multiset $T'$ has size
\[
  |T'|\le |T|\frac{1/\varepsilon}{\varepsilon/(\alpha|T|)} + |T|
  =\frac{\alpha|T|^2}{\varepsilon^2} + |T|.
\]

Apply the $\alpha$-approximation to the Steiner tree version of the Cable-Trench Problem
with terminal set $T'$, $\lambda_1=1$, and
$\lambda_2=\varepsilon/(\alpha|T|)$. In the resulting tree, merge copies at the same
vertex and add the direct metric edge from the root to every heavy terminal. Take a shortest-path
tree $A'$ of the resulting graph. We show that $A'$ satisfies the desired approximation guarantee for the original
Uniform Cost-Distance instance.

Write
$\OPT_{\mathrm{CT}}$ for the optimum value of the Cable-Trench instance and
$\OPT_{\mathrm{CD}}$ for the optimum value of the original instance. Let
\[
  D_{\mathrm{heavy}}:=\sum_{t\text{ heavy}}w(t)c(r,t).
\]
Since $w(t)\ge1/\varepsilon$ for a heavy terminal, the direct edge to $t$ costs at most an
$\varepsilon$ fraction of its minimum delay contribution. Hence
\[
  \cost(A')\le(1+\varepsilon)D_{\mathrm{heavy}}+\alpha\OPT_{\mathrm{CT}}.
\]

To upper-bound the Cable-Trench optimum, evaluate an optimal Uniform Cost-Distance solution for the
original problem under the Cable-Trench objective: ignore the heavy terminals and use the
rounded weights for all other terminals. The weight of each remaining terminal increases by at most
$\varepsilon/(\alpha|T|)$. Every root-to-terminal path in the optimal tree has length at most the tree's connection
cost, and hence at most $\OPT_{\mathrm{CD}}-D$, where $D=\sum_t w(t)c(r,t)$. Therefore
\begin{align*}
  \OPT_{\mathrm{CT}}
  &\le \OPT_{\mathrm{CD}}-D_{\mathrm{heavy}}
     +|T|\frac{\varepsilon}{\alpha|T|}(\OPT_{\mathrm{CD}}-D)\\
  &\le \left(1+\frac{\varepsilon}{\alpha}\right)
       (\OPT_{\mathrm{CD}}-D_{\mathrm{heavy}}).
\end{align*}
Because $\alpha\ge1$,
\[
\begin{aligned}
  \cost(A')
  &\le(1+\varepsilon)D_{\mathrm{heavy}}+\alpha\OPT_{\mathrm{CT}}\\
  &\le(1+\varepsilon)D_{\mathrm{heavy}}
    +(\alpha+\varepsilon)(\OPT_{\mathrm{CD}}-D_{\mathrm{heavy}})
  \le(\alpha+\varepsilon)\OPT_{\mathrm{CD}}.
\end{aligned}
\]
\end{proof}

\section{Conclusion}

We introduced an anchored-gate algorithm that compares direct reconnection
at capacity $\mu$ with pairwise reconnection at half that capacity.
The complementary transportation bounds and charging the
matching-cost to an optimal tree allow the analysis to beat the
decomposition gap.
The resulting approximation factors are $1.698$ for spanning instances
and $2.041$ for general terminal sets, with deterministic versions
obtained by enumerating the relevant offsets.

Staircase instances yield lower bounds approaching $1+1/\sqrt2$ for
cut-and-reconnect algorithms that start from a minimum spanning tree and $3/2$ for
the algorithm presented here.

The reduction shows that any improved approximation guarantee for the Steiner version of the
Cable-Trench Problem, in which all delay weights are equal, transfers to the general weighted
problem up to an arbitrarily small additive loss.

Extending the pairwise reconnection framework to $k$-component reconnection appears
difficult because grouping $k$ components is no longer a standard graph-matching problem.
Further work could seek to
determine the true approximation ratio of the present algorithm and whether it is optimal
within the class of pairwise-reconnection algorithms.

\section*{Use of generative AI}

Before using generative AI, the author had developed an improved approximation algorithm based on component splitting that beats the decomposition gap by some small $\varepsilon > 0$, as part of the author's master's thesis~\cite{Perner2025Thesis}. The present work was initially intended to publish the results of this thesis. During the preparation of the paper, the thesis was provided to ChatGPT 5.6 Sol. The model suggested replacing the component-splitting approach by Foos, Held, and Spitzley~\cite{FoosHeldSpitzley2023Conference} with a formulation similar to the gate-based approach presented in this paper. Combining this formulation with the ideas and analysis from the thesis leads to a simpler analysis and improves the gain in the approximation guarantee to the numbers presented here.

The two-candidate analysis, the pairwise matching construction, the matching-cost bound based on edge-disjoint paths, the lower-bound examples, and the problem reduction were developed by the author.

The model was also used to assist with restructuring, language editing, and the derivation of inequality bounds. All AI-generated mathematical ideas and arguments were independently reviewed and verified by the author.

\newpage
\appendix
\renewcommand*{\theHsection}{appendix.\arabic{section}}
\renewcommand*{\theHsubsection}{\theHsection.\arabic{subsection}}

\section{Bounding the supremum for the approximation ratio}\label{app:factor-proof}

This appendix certifies the numerical guarantees stated in
\Cref{sec:contributions}. Throughout, $\omega\ge1$ and $\lambda>0$.

\subsection{Eliminating the optimum-value parameter}

Fix $(x,u)\in(0,1)\times[0,1]$ and write
\[
  A:=A_{\omega,\lambda}(x,u),\qquad
  B:=B_{\omega,\lambda}(x,u),\qquad
  h:=h_{\omega,\lambda}(x).
\]
By \eqref{eq:general-factor-definition}, the dependence on
$p\ge1$ is through
\[
  \frac{A}{p},
  \qquad
  1+h+\frac{B-1-h}{p}.
\]
If $B\ge1+h$, both expressions are nonincreasing in $p$,
and their minimum is maximized at $p=1$.
If $B<1+h$, the first expression is decreasing and
the second is increasing. Hence their minimum is
maximized at $p=1$ or, if it lies in $[1,\infty)$,
at their intersection. Solving for $p$ in the
equality case gives
\[
  p=1+\frac{A-B}{1+h}.
\]
Thus the intersection lies in $[1,\infty)$ exactly
when $A\ge B$, and the value at the intersection is
\[
  I_{\omega,\lambda}(x,u):=
  \frac{A(1+h)}{A-B+1+h}.
\]
Therefore, \eqref{eq:general-factor-definition}
reduces to maximizing, over
$(x,u)\in(0,1)\times[0,1]$,
\begin{equation}\label{eq:reduced-factor}
  Q_{\omega,\lambda}(x,u):=
  \begin{cases}
    \max\{\min\{A,B\},I_{\omega,\lambda}(x,u)\},
      & \substack{B<1+h\\ A\ge B},\\[2mm]
    \min\{A,B\},
      & \text{otherwise}.
  \end{cases}
\end{equation}

\subsection{Reducing to one dimension}
\label{sec:one-dimensional-certification}

We now reduce the maximization over $(x,u)$ to three one-dimensional terms.
For $x\in(0,1)$, recall
\begin{equation*}
  g_\omega(x):=\sqrt{2\omega x(1-x)},
  \qquad
  h_{\omega,\lambda}(x):=
  \frac1\lambda\sqrt{\frac{\omega x}{2(1-x)}}.
\end{equation*}
The definitions of $A_{\omega,\lambda}$ and $B_{\omega,\lambda}$ can then be
written as
\begin{align}
 A_{\omega,\lambda}(x,u)
 &=1+(\omega-1)x+\frac{g_\omega(x)}{2\lambda}
   +\frac{\lambda g_\omega(x)}{2}(1-u),
 \label{eq:A-u-compact}\\
 B_{\omega,\lambda}(x,u)
 &=1+\left(\omega-\frac12\right)x+\frac{g_\omega(x)}{2\lambda}
   +\lambda g_\omega(x)(\sqrt{u}-u).
 \label{eq:B-u-compact}
\end{align}
In particular,
\begin{equation}\label{eq:AB-difference-u}
 A_{\omega,\lambda}(x,u)-B_{\omega,\lambda}(x,u)
 =\frac{\lambda g_\omega(x)}{2}
   \left((1-\sqrt{u})^2-\frac{x}{\lambda g_\omega(x)}\right),
\end{equation}
and
\begin{equation}\label{eq:g-h-identity}
  \frac{g_\omega(x)}{2\lambda}=(1-x)h_{\omega,\lambda}(x),
  \qquad
  h_{\omega,\lambda}(x)=\frac{\omega x}{\lambda g_\omega(x)}.
\end{equation}

\begin{lemma}[The intersection term is harmless]
\label[lemma]{lem:intersection-harmless-compact}
For every fixed $x\in(0,1)$,
\[
 \sup_{0\le u\le1}Q_{\omega,\lambda}(x,u)
 =
 \sup_{0\le u\le1}
 \min\{A_{\omega,\lambda}(x,u),B_{\omega,\lambda}(x,u)\}.
\]
\end{lemma}

\begin{proof}
Fix $x\in(0,1)$. By \eqref{eq:reduced-factor}, it suffices
to bound $I_{\omega,\lambda}(x,u)$ on the active region
\begin{equation}\label{eq:active-region}
 A_{\omega,\lambda}(x,u)\ge B_{\omega,\lambda}(x,u),
 \qquad B_{\omega,\lambda}(x,u)<1+h_{\omega,\lambda}(x).
\end{equation}
If this region is empty, there is nothing to prove.

\emph{Step 1: Determine the relevant region.}
Since $B_{\omega,\lambda}(x,u)\ge B_{\omega,\lambda}(x,0)$
by \eqref{eq:B-u-compact}, the second condition in
\eqref{eq:active-region} and \eqref{eq:g-h-identity} give
\[
 0<1+h_{\omega,\lambda}(x)-B_{\omega,\lambda}(x,0)
 =x\left(h_{\omega,\lambda}(x)-\omega+\frac12\right).
\]
Using $h_{\omega,\lambda}(x)=\omega x/(\lambda g_\omega(x))$,
we obtain
\begin{equation}\label{eq:x-over-lambda-g-lower}
 \frac{x}{\lambda g_\omega(x)}
 >1-\frac{1}{2\omega}\ge\frac12.
\end{equation}
On the other hand, the first condition in
\eqref{eq:active-region} and \eqref{eq:AB-difference-u}, together with
\eqref{eq:x-over-lambda-g-lower}, imply $(1-\sqrt{u})^2\ge x/(\lambda g_\omega(x))$.
Consequently,
\begin{equation}\label{eq:sqrt-u-small}
 0\le\sqrt{u}
 \le1-\sqrt{\frac{x}{\lambda g_\omega(x)}}
 <1-\frac1{\sqrt2}<\frac13.
\end{equation}
The first two inequalities also imply $x/(\lambda g_\omega(x))\le1$,
so $h_{\omega,\lambda}(x)\le\omega$ by
\eqref{eq:g-h-identity}.

\emph{Step 2: Bound the intersection term by the arithmetic mean.}
We start as in Step 1, now using $h_{\omega,\lambda}(x)\le\omega$ to get
\begin{align*}
 0&<1+h_{\omega,\lambda}(x)-B_{\omega,\lambda}(x,0) = x\left(h_{\omega,\lambda}(x)-\omega+\frac12\right) \le\frac{x}{2}<1\le A_{\omega,\lambda}(x,u),
\end{align*}
where the last inequality follows from
\eqref{eq:A-u-compact}. Hence, by the definition
of $I_{\omega,\lambda}$,
\begin{align*}
 &I_{\omega,\lambda}(x,u)-B_{\omega,\lambda}(x,u)\\
 &=\frac{(A_{\omega,\lambda}(x,u)-B_{\omega,\lambda}(x,u))
 (1+h_{\omega,\lambda}(x)-B_{\omega,\lambda}(x,u))}
 {A_{\omega,\lambda}(x,u)+1+h_{\omega,\lambda}(x)-B_{\omega,\lambda}(x,u)}\\
 &\le\frac{A_{\omega,\lambda}(x,u)-B_{\omega,\lambda}(x,u)}2.
\end{align*}
Thus
\begin{equation}\label{eq:intersection-mean-bound}
 I_{\omega,\lambda}(x,u)
 \le\frac{A_{\omega,\lambda}(x,u)+B_{\omega,\lambda}(x,u)}2.
\end{equation}

\emph{Step 3: Compare with the crossing.}
By \eqref{eq:AB-difference-u} the crossing point of $A_{\omega,\lambda}(x,u)$ and $B_{\omega,\lambda}(x,u)$ is
\[
 \sqrt{u_*}:=1-\sqrt{\frac{x}{\lambda g_\omega(x)}}.
\]

Moreover, \eqref{eq:sqrt-u-small} gives
$0\le\sqrt{u}\le\sqrt{u_*}<1/3$ on the relevant region.

Differentiating \eqref{eq:A-u-compact} and
\eqref{eq:B-u-compact} with respect to $\sqrt{u}$ yields
\[
 \frac{d}{d\sqrt{u}}
 \frac{A_{\omega,\lambda}(x,u)+B_{\omega,\lambda}(x,u)}2
 =\frac{\lambda g_\omega(x)}2(1-3\sqrt{u})>0
\]
for $0\le\sqrt{u}\le\sqrt{u_*}$.
Together with \eqref{eq:intersection-mean-bound}, this gives
\begin{align*}
 I_{\omega,\lambda}(x,u)
 &\le\frac{A_{\omega,\lambda}(x,u)+B_{\omega,\lambda}(x,u)}2\\
 &\le\frac{A_{\omega,\lambda}(x,u_*)+B_{\omega,\lambda}(x,u_*)}2\\
 &=\min\{A_{\omega,\lambda}(x,u_*),B_{\omega,\lambda}(x,u_*)\}\\
 &\le\sup_{0\le v\le1}
 \min\{A_{\omega,\lambda}(x,v),B_{\omega,\lambda}(x,v)\}.
\end{align*}
Thus the intersection term cannot increase the supremum
in \eqref{eq:reduced-factor}, proving the claim.
\end{proof}

\begin{lemma}[Three candidates]
\label[lemma]{lem:three-candidates-compact}
For fixed $x\in(0,1)$, the maximum of
\[
 \min\{A_{\omega,\lambda}(x,u),B_{\omega,\lambda}(x,u)\}
\]
over $u\in[0,1]$ is attained at $u=0$, at $u=1/4$, or at the crossing
\begin{equation}\label{eq:crossing-u}
 \sqrt{u}=1-\sqrt{\frac{x}{\lambda g_\omega(x)}},
\end{equation}
provided the right-hand side lies in $[0,1/2]$. At the crossing, the common
value equals
\begin{equation}\label{eq:F-compact}
 F_{\omega,\lambda}(x)
 =1+\left(\omega-\frac32\right)x
   +\frac{g_\omega(x)}{2\lambda}
   +\sqrt{\lambda g_\omega(x)x}.
\end{equation}
\end{lemma}

\begin{proof}
As $\sqrt{u}$ increases, \eqref{eq:A-u-compact} shows that
$A_{\omega,\lambda}(x,u)$ is nonincreasing, while
\eqref{eq:B-u-compact} shows that $B_{\omega,\lambda}(x,u)$ is nondecreasing
for $\sqrt{u}\in[0,1/2]$ and nonincreasing for
$\sqrt{u}\in[1/2,1]$. Hence their minimum is nonincreasing on
$[1/2,1]$. On $[0,1/2]$, its maximum can occur only at an endpoint or where
the two functions cross. By \eqref{eq:AB-difference-u}, there is at most one
crossing, and it is given by \eqref{eq:crossing-u}. Substituting
\eqref{eq:crossing-u} into \eqref{eq:A-u-compact} yields
\eqref{eq:F-compact}.
\end{proof}

By the two lemmas, the candidate at $u=0$ is at most
$B_{\omega,\lambda}(x,0)$, the candidate at $u=1/4$ is at most
$A_{\omega,\lambda}(x,1/4)$, and the crossing candidate equals
$F_{\omega,\lambda}(x)$. Therefore
\begin{equation}\label{eq:three-one-dimensional-bounds}
 \mathcal R_{\omega,\lambda}
 \le
 \max\left\{
   \sup_{0\le x\le1}B_{\omega,\lambda}(x,0),
   \sup_{0\le x\le1}A_{\omega,\lambda}\!\left(x,\frac14\right),
   \sup_{0\le x\le1}F_{\omega,\lambda}(x)
 \right\}.
\end{equation}
Here we have dropped the restriction in \eqref{eq:crossing-u}, which can only
increase the upper bound. We now show how to evaluate the three suprema independently.

For $a\in\mathbb R$ and $b\ge0$, define
\[
  M(a,b):=1+\frac{a+\sqrt{a^2+b^2}}{2}.
\]

\begin{lemma}[Maximum of a one-variable expression]
\label[lemma]{lem:semicircle-maximum}
For every $a\in\mathbb R$ and $b\ge0$, the maximum of
$1+ax+b\sqrt{x(1-x)}$ on $[0,1]$ is attained, and
\begin{equation}\label{eq:semicircle-compact}
  \max_{0\le x\le1}
  \left\{1+ax+b\sqrt{x(1-x)}\right\}=M(a,b).
\end{equation}
\end{lemma}

\begin{proof}
Making the substitution $x=(1+y)/2$, with $y\in[-1,1]$,
gives
\[
  1+ax+b\sqrt{x(1-x)}
  =1+\frac a2+\frac12\left(ay+b\sqrt{1-y^2}\right).
\]
By the Cauchy--Schwarz inequality,
\[
  ay+b\sqrt{1-y^2}
  \le \sqrt{a^2+b^2}\,
      \sqrt{y^2+(1-y^2)}
  =\sqrt{a^2+b^2},
\]
proving the claim.
\end{proof}

Applying \eqref{eq:semicircle-compact} gives
\begin{align}
  \sup_{0\le x\le1} B_{\omega,\lambda}(x,0)
  &=M\left(\omega-\frac12,
      \frac{\sqrt{2\omega}}{2\lambda}\right),
  \label{eq:B-sup-evaluation}\\
  \sup_{0\le x\le1} A_{\omega,\lambda}\!\left(x,\frac14\right)
  &=M\left(\omega-1,
      \sqrt{2\omega}\left(\frac{1}{2\lambda}
      +\frac{3\lambda}{8}\right)\right).
  \label{eq:A-sup-evaluation}
\end{align}

For the crossing term,
\[
 \sqrt{\lambda g_\omega(x)x}
 =\sqrt{\lambda}(2\omega)^{1/4}
   x^{3/4}(1-x)^{1/4}.
\]
For every $\eta>0$, applying $2\sqrt{ab}\le\eta a+b/\eta$ with
$a=x$ and $b=\sqrt{x(1-x)}$ yields
\begin{align*}
 F_{\omega,\lambda}(x)
 &\le 1+a_Fx+b_F\sqrt{x(1-x)},\\
 a_F
 &:=\omega-\frac32
   +\frac{\eta}{2}\sqrt{\lambda}(2\omega)^{1/4},\\
 b_F
 &:=\frac{\sqrt{2\omega}}{2\lambda}
   +\frac{\sqrt{\lambda}(2\omega)^{1/4}}{2\eta}.
\end{align*}
Consequently,
\begin{equation}\label{eq:F-sup-evaluation}
  \sup_{0\le x\le1}F_{\omega,\lambda}(x)\le M(a_F,b_F).
\end{equation}

For spanning instances we take $\omega=1$. For general terminal sets we take
$\omega=13863/10000$, which is admissible because
$\ln(4)<13863/10000$.
Substitution into \eqref{eq:B-sup-evaluation},
\eqref{eq:A-sup-evaluation}, and \eqref{eq:F-sup-evaluation}
gives the following bounds.

\begin{center}
\begin{tabular}{cccccc}
\toprule
$\omega$ & $\lambda$ & $\eta$
& $\sup_x B_{\omega,\lambda}(x,0)$
& $\sup_x A_{\omega,\lambda}(x,1/4)$
& \multicolumn{1}{c}{$\sup_x F_{\omega,\lambda}(x)$} \\
\midrule
$1$ & $11/10$ & $25/26$
& $<1.657193$ & $<1.613094$ & $<1.697437$ \\
$13863/10000$ & $43/40$ & $19/22$
& $<2.031651$ & $<1.941370$ & $<2.040546$ \\
\bottomrule
\end{tabular}
\end{center}

All parameters in the table are rational, and the bounds follow from formulas
containing only arithmetic operations and square roots. Each displayed
strict inequality can be verified using directed rational interval arithmetic with 30 decimal digits of precision at each step.

Combining the table with \eqref{eq:three-one-dimensional-bounds} yields
\[
 \mathcal R_{1,11/10}<1.698,
 \qquad
 \mathcal R_{13863/10000,43/40}<2.041.
\]
This proves the numerical approximation guarantees stated in
\Cref{sec:contributions}.

\end{document}